\documentclass[prd,twocolumn]{revtex4}
\usepackage{graphicx}
\usepackage{amsmath}
\usepackage{amssymb}
\usepackage{natbib}
\newcommand{\mat}[1]{\mathbf{#1}}
\newcommand{\argmin}{\operatornamewithlimits{arg\ min}}
\begin{document}
\title{Spline Based Search Method For Unmodeled Transient Gravitational Wave Chirps}
\author{Soumya D.~Mohanty}
\affiliation{Department of Physics and Astronomy, The University of Texas Rio Grande Valley, One West University Blvd.,
Brownsville, Texas 78520, USA}
\date{April 2017}

\begin{abstract}
A method is described for the detection and estimation of transient chirp signals that 
are characterized by smoothly evolving,
but otherwise unmodeled, amplitude envelopes and instantaneous frequencies. Such
signals are particularly relevant for gravitational wave searches, where they may arise in a wide range of astrophysical scenarios.
The method uses 
splines with continuously adjustable breakpoints to represent the amplitude envelope and instantaneous
frequency of a signal, and estimates them from noisy data using penalized least squares
and model selection. 
Simulations based on waveforms spanning a wide morphological range show that the method  performs well in
a signal-to-noise ratio regime 
where the time-frequency signature of a signal is highly degraded, thereby extending the coverage of current 
unmodeled gravitational wave searches to a wider class of signals. 
\end{abstract}
\maketitle

\section{Introduction}
\label{intro}
The tally of confirmed direct gravitational wave (GW) detections now 
stands at 5 events.
Across  two observing runs,
the twin Advanced Laser Interferometer Gravitational-Wave Observatory (LIGO)~\cite{2016PhRvL.116m1103A} detectors found
 GW150914~\cite{PhysRevLett.116.061102}, GW151226~\cite{PhysRevLett.116.241103}, and GW170104~\cite{PhysRevLett.118.221101} in two-way coincidence. All three signals
 are consistent with binary black hole (BBH) mergers.
(A fourth BBH merger,
LVT151012, was discovered in the first observing run but 
 with marginal significance.) 
These were followed by an additional BBH merger signal,
GW170814~\cite{abbott2017gw170814}, that was also the first event to be 
discovered in a three-way coincidence between
the LIGO and Advanced Virgo~\cite{0264-9381-32-2-024001} detectors. 
The second observing run concluded with the remarkable 
discovery by LIGO and Virgo of GW170817~\cite{PhysRevLett.119.161101},  a binary neutron star inspiral. 
 
Along with further improvements in the sensitivity of 
the LIGO and Virgo detectors,  additional second
 generation detectors -- KAGRA~\cite{2012CQGra..29l4007S} and 
 LIGO-India~\cite{2013IJMPD..2241010U} -- are scheduled to come online over the next several years. Besides
 significantly enhancing overall search sensitivity, combining the data from the upcoming
 network of  detectors will  better 
 localize sources on the sky, thereby increasing the chances of finding their electromagnetic counterparts.

The type of signals detected by LIGO so far have waveforms that 
 can be calculated theoretically as a function of system parameters,  allowing parametric data analysis methods, such
as matched filtering~\cite{Helstrom}, to be used for their detection and estimation. This approach is inapplicable, however, to 
astrophysical sources that are unanticipated or that emit inherently unpredictable
signals. Search methods for such unmodeled signals -- known as GW bursts when they are
transient -- must use minimal prior assumptions about their waveforms. 

Due to their 
broad scope, burst search 
methods can also detect sufficiently strong parametric signals.
In fact, GW150914 was first detected~\cite{PhysRevD.93.122004} by a burst search method~\cite{PhysRevD.93.042004}
that implements a regularized maximum likelihood analysis~\cite{klimenko:2005,2006CQGra..23.4799M,2006CQGra..23S.673R} of data 
from a network of 
detectors. 
Refined estimates of the source parameters, such as the component masses, were 
obtained subsequently using parametric methods.

Among burst signals, the most challenging to search for are those that do 
not have compact time-frequency signatures. 
We know of several possible astrophysical 
scenarios where such signals may arise.  
Among these
are (i) the post core-bounce phase of a core-collapse
supernova (CCSN)\cite{2006PhRvL..96t1102O,ott2009gravitational}, (ii) dynamical
instabilities in rotating newborn neutron stars~\cite{2002PhRvD..65l4003L,2012ApJ...761...63P}, and (iii) clump formation
or 
dynamical instabilities  
in the accretion disc surrounding a newly formed black hole  
in a collapsar~\cite{PhysRevLett.87.091101,PhysRevLett.106.251102}.  
Such signals may be generic to
GW emissions powered by the rotational energy of a
compact engine.

In many of these scenarios, the burst signals 
spread their total energy over well defined ``tracks"  in the time-frequency plane. 
Such signals are
generally called {\em chirps} in the signal processing literature. (BBH merger
signals are examples of parameteric chirps.) 
Taking the analytic representation, $a(t) \exp(i \phi(t))$, of 
a signal $s(t)$, one expects a 
track-like feature
in the time-frequency representation of $s(t)$ when the amplitude envelope
$a(t)$ and instantaneous frequency $f(t) = \dot{\phi}(t)$ evolve adiabatically -- $f(t) \gg \dot{a}/a$ and $f^2(t) \gg \dot{f}(t)$ -- 
relative to the instantaneous period $1/f(t)$. 
The sharpness of the track is determined by $\dot{a}(t)$, with a smaller $\dot{a}$ leading to a sharper track. 

Several search methods have been developed in the
GW data analysis literature for short duration ($\sim 1$ sec)
unmodeled chirps for which $a(t)$, or $f(t)$, or both are unknown.
 The Track-Search method~\cite{1999PhRvD..60j2001A} uses an image processing
 approach to search for track-like features in the 
 Wigner-Ville (WV) time-frequency distribution~\cite{cohen1989time}.
 This method can detect signals with arbitrary 
$a(t)$ and $f(t)$ provided its track is distinguishable from the 
spurious features that appear in the WV transform due to its non-linear
nature. A significant advance has recently been made in 
mitigating these spurious features by applying sparsity 
regularization to the WV transform~\cite{addesso2015sparsifying} but it remains
to be integrated with methods such as Track-Search. 

A natural approach to the detection of unmodeled chirps is to approximate 
$f(t)$ by a piecewise linear curve. Each 
piece represents a transient linear
chirp signal, called a chirplet, and the signal is assumed to be 
a sequence of connected chirplets called a chirplet chain. 
(Note that a chirplet chain only approximates $f(t)$ and that 
additional degrees of freedom are needed to model $a(t)$.)
The main challenge in this approach is the extremely high computational 
cost of  searching
the space of all chirplet chains to find the one that best fits
the data. Different methods have been proposed to address this issue. 

The Best Chirplet Chain 
method~\cite{chassande2006best} approximates the chirplet chain approach, for the 
special case where the  
duration (or scale) of chirplets  and the length of a chain (number of chirplets)
are fixed, 
with path integrals of the WV transform.
The Chirplet Path Pursuit (CPP) method~\cite{candes2008gravitational}, 
uses a different approach in which the chains are constructed out of
a discrete set of multiscale chirplets. 
The use of multiscale chirplets allows greater flexibility
in the signal model since slowly evolving parts of $f(t)$
can be approximated by longer chirplets. 
This leads to chains with variable lengths and 
a selection of the best fit chirplet chain requires balancing the fitness of a chain
against its complexity  using a penalty on the chain length. 

Track detection~\cite{thrane2011long} or path integrals~\cite{thrane2014seedless} have also
been studied
for detecting long duration ($\gtrapprox 10$~sec) GW 
bursts in data from multiple detectors.  Unlike the single detector methods
mentioned above, these methods use the 
short-time cross-spectra of data from pairs of GW detectors. In the following, 
we consider only the single detector case, leaving multiple detectors to future work.

While the chirplet chain approach is designed to approximate $f(t)$, it is important
to consider $a(t)$ also. In particular, 
there is no reason for $a(t)$ to remain constant over the 
duration of a single chirplet. This issue was addressed in CPP
 by allowing the amplitude of each chirplet to evolve as a polynomial in time.
However, the degree of the polynomial is fixed for all the chirplets at a given scale
and has to be prescribed in 
advance. This is difficult to do when $a(t)$ has an unknown and complex evolution.

In this paper, we present a search method for chirp signals that 
explicitly takes amplitude modulation into account.
The structure of the method   
follows logically from modeling $a(t)$ and $f(t)$ as
independent splines
and seeking a computationally feasible solution 
to the resulting high-dimensional non-linear regression problem. 
Early and intermediate
steps in the development of 
the method were reported in~\cite{mohanty2012particle,2012AstRv...7d...4M,mohanty2014}.

 The detection and estimation performance of the method is 
 quantified using simulated data, incorporating a wide range 
 of signal waveform morphologies, in a signal-to-noise ratio (SNR) regime
 where the signal track in the time-frequency plane is 
easily disrupted and masked by noise. 
To keep computational costs under control, we focus only on 
signals with durations of $O(1)$~sec in this paper although
the general idea
can, in principle, be applied to much longer signals. 

 The rest of the paper is organized as follows. Sec.~\ref{model}
sets up the notation and the models used in this paper for 
noise and signal.
Sec.~\ref{seecr} presents a description of the method. 
The simulation set up  used in assessing its 
detection and estimation performance  are described in Sec.~\ref{sims}. The results obtained from 
 the simulations are presented in Sec.~\ref{results}.
  Sec.~\ref{seecr_tf_detestperf} compares the performance of the method  with that of time-frequency clustering, a key component of the burst search methods currently used in LIGO. 
 This is followed, in Sec.~\ref{seecr_cpp_trksrch_detestperf},
 by comparisons with Track-Search and  CPP.
 Sec.~\ref{conclusions} presents the conclusions from our study.
 
 \section{Statistical Model}
 \label{model}

In the following, a symbol such as $\overline{s}\in \mathbb{R}^N$ denotes a row vector with $N$ elements,
and $s_j$, $j = 0,1,\ldots,N-1$, or $[\overline{s}]_j$,
denotes its $j^{\rm th}$ element.
When $\overline{s}$ is a finite length discrete-time sequence of
sample values  of an underlying continuous-time function $s(t)$, the sampling times are denoted by $t_i$, $i = 0, 1, \ldots, N-1$, and $s_i = s(t_i)$.
A symbol such as $\widehat{s}$ denotes either a solution to an optimization problem or a quantity estimated from data. 

Boldface symbols, such as $\mat{A}$,
 denote matrices 
with the 
element in its $i^{\rm th}$ row and $j^{\rm th}$ column denoted by $A_{ij}$ or $[{\bf A}]_{ij}$. The identity matrix is denoted by $\mat{I}$. 

We use $\widetilde{s}$ to denote the Discrete Fourier Transform (DFT) of $\overline{s}$,
\begin{eqnarray}
\widetilde{s}^T & := & \mat{F} \overline{s}^T\;,\label{dft}\\
F_{km} & = & e^{-2\pi i km /N}\;, 
\end{eqnarray}
with $\widetilde{s}_j$ being its $j^{\rm th}$
element.
The inverse DFT is given by,
\begin{eqnarray}
\mat{F}^{-1} & = & \frac{1}{N}\mat{F}^\dagger\;.\label{invF}
\end{eqnarray}
The symbol  `$./$' denotes element-by-element division, and
the supremum of integers less than or equal to $x \in \mathbb{R}_{>0}$ is denoted by
$\left\lfloor x \right\rfloor$.


\subsection{Noise Model}
\label{noisemodel}

We will denote a segment of  GW detector output, sampled uniformly with a sampling frequency $f_s$, by $\overline{y}\in \mathbb{R}^N$. Under the null and alternative hypotheses, denoted by $H_0$ and $H_1$
respectively,
\begin{eqnarray}
\overline{y} & = & \left\{\begin{array}{lc}
    \overline{n} &  \;; H_0\\
   \overline{s}+\overline{n}  & \;; H_1
\end{array}\right.\;,
\label{datamodel}
\end{eqnarray}
where $\overline{s}$ is a GW signal and $\overline{n}$ is a realization of noise.
Our noise model assumes that $\overline{n}$ is drawn from 
a zero mean, Gaussian, stationary stochastic process. 
Let $\mat{C}$, $C_{ij} = E[n_i n_j]$, 
where $E[A]$ denotes the ensemble average of a random variable $A$, be the covariance matrix of the noise
segment.

Since $\mat{C}$ is symmetric and positive definite, an inner product
can be defined on $\mathbb{R}^N$,
\begin{eqnarray}
\langle \overline{x}, \overline{y} \rangle & = & \overline{x}\mat{C}^{-1}\overline{y}^T\;.
\label{innprod}
\end{eqnarray}
The norm induced by this inner product will be denoted by $\|\overline{x}\|^2 = \langle \overline{x},\overline{x}\rangle$.
It can be shown that
\begin{eqnarray}
\|\overline{x}\|^2 & = & \frac{1}{N} \widetilde{x}^\ast \left(\mat{F}\mat{C}
\mat{F}^{-1}\right)^{-1}\widetilde{x}^T\nonumber\\
&\approx & 
\frac{1}{N f_s}\widetilde{x}\left(\widetilde{x}^\dagger ./\overline{S}^T\right)\;,
\label{norm}
\end{eqnarray}
where $\overline{S}$ is the two-sided power spectral density (PSD) of the noise defined by
\begin{eqnarray}
S_i & = & \frac{1}{N f_s}E\left[|\widetilde{n}_i|^2\right] = \frac{1}{f_s}\left(
\mat{F}\mat{C}\mat{F}^{-1}
\right)_{ii}\;.
\end{eqnarray}
It follows that,
\begin{eqnarray}
\delta_f\sum_{m=0}^{N-1} S_m & = & \sigma^2\;,
\end{eqnarray}
where $\sigma^2$ is the variance of the noise and $\delta_f = f_s/N$ is the 
spacing between consecutive frequencies in the DFT.

The approximation in Eq.~(\ref{norm})
arises from neglecting off-diagonal terms in $\mat{F}\mat{C}
\mat{F}^{-1}$. However, the approximation approaches
equality very rapidly with an increase in $N$.

\subsection{Signal Model}
\label{signalmodel}
 As discussed earlier,  the
 amplitude envelope $a(t)$ and instantaneous frequency $f(t) = \dot{\phi}(t)$ of
 a chirp signal evolve smoothly on the timescale of the instantaneous
 period $1/f(t)$. We model
 this smoothness behavior by prescribing $a(t)$ and $f(t)$ to be splines. 
 The motivation behind using splines in particular is discussed 
 further in Sec.~\ref{seecr}. Appendix~\ref{app_bsplines} provides a brief review of splines and B-spline functions.  
 
 Let $a(t;\overline{\alpha},\overline{\tau}_a)$ denote the spline for $a(t)$, where $\overline{\tau}_a$ are the breakpoints, 
 \begin{eqnarray}
 a(t;\overline{\alpha},\overline{\tau}_a) & = & \sum_{j=0}^{M-1} \alpha_j \mathcal{B}_{j,k}(t;\overline{\tau}_a)\;,
 \label{aoft_spline}
 \end{eqnarray}
 and $\mathcal{B}_{j,k}(t;\overline{\tau}_a)$ is a B-spline function~\cite{deBoor} 
 of order $k$.
Since B-splines have compact support,  $a(t;\overline{\alpha},\overline{\tau}_a)  =  0$ for $t\notin \left[\tau_{a,0},\tau_{a,M-1}\;\right]$, where $\tau_{a,i} = [\overline{\tau}_a]_i$.
 As we will see later, the linear dependence of $a(t;\overline{\alpha},\overline{\tau}_a)$ on $\overline{\alpha}$ allows
 considerable simplification in the analysis.

 Let $f(t;\overline{\nu},\overline{\tau}_f)$ be the spline corresponding to $f(t)$, where  $\overline{\tau}_f\in \mathbb{R}^K$  and $\overline{\nu}\in \mathbb{R}^K$ denote the breakpoints and corresponding instantaneous frequencies 
 that the spline must interpolate.
 Unlike $a(t;\overline{\alpha},\overline{\tau}_a)$, there is no particular advantage gained by 
 expressing $f(t;\overline{\nu},\overline{\tau}_f)$ in terms of B-spline functions. 
 We use Steffen's method~\cite{1990A&A...239..443S} for spline interpolation, which 
 guarantees the monotonicity of the interpolating function between given data points,
 in order to prevent spurious oscillations in $f(t;\overline{\nu},\overline{\tau}_f)$. 
 
With   $\overline{\nu}$, $\overline{\tau}_a$, and $\overline{\tau}_f$ denoted collectively by $\overline{\theta}$, the signal model is given by,
\begin{eqnarray}
s(t_i;\overline{\alpha},\overline{\theta},\phi_0) & = &
a(t_i;\overline{\alpha},\overline{\tau}_a) \sin(\phi(t_i;\overline{\nu},\overline{\tau}_f)+\phi_0)
\label{seecrsigmodel}\;,\\
\phi(t;\overline{\nu},\overline{\tau}_f) & = & \left\{ \begin{array}{ll}
0, & t < \tau_{a,0}\\
\int_{\tau_{a,0}}^t dt^\prime f(t^\prime ;\overline{\nu},\overline{\tau}_f), & t\leq \tau_{a,M-1}
\end{array}
\right.
\end{eqnarray}
Let $\mat{X}_0$ and $\mat{X}_1$ denote matrices
given by
\begin{eqnarray}
[\mat{X}_0(\overline{\theta})]_{jm} & = & \mathcal{B}_{j,k}(t_m;(\overline{\tau}_a)
\sin(\phi(t_m;\overline{\nu},\overline{\tau}_f))\;,
\end{eqnarray}
and
\begin{eqnarray}
[\mat{X}_1(\overline{\theta})]_{jm} & = & \mathcal{B}_{j,k}(t_m;\overline{\tau}_a)
\cos(\phi(t_m;\overline{\nu},\overline{\tau}_f))\;.
\end{eqnarray}
In terms of these matrices, the signal sequence is,
\begin{eqnarray}
\overline{s}(\overline{\alpha},\overline{\theta},\phi_0)
& = &  \overline{\beta}{\bf X}(\overline{\theta})\;,
\label{seecrsigvec}\\
\overline{\beta} & = & \overline{\alpha}\mat{\Phi}_0\;,\nonumber\\
\mat{\Phi}_0 & = & \left(\begin{array}{cc}
 \cos\phi_0\mat{I} & \sin\phi_0\mat{I} 
\end{array}\right)\;,\\
{\bf X}(\overline{\theta}) & = & \left(\begin{array}{c}
{\bf X}_0(\overline{\theta})\\
{\bf X}_1(\overline{\theta})
\end{array}
\right)\;.
\label{xmat}
\end{eqnarray}
While the signal
model in Eq.~(\ref{seecrsigmodel})
captures the basic idea of smoothness in the evolution of $a(t)$ and $f(t)$, 
it does not enforce the adiabaticity requirement. This is mainly because 
it is technically difficult to incorporate this constraint at present. As a result,
the scope of the model actually encompasses a broader set of signals than just 
well-defined chirps.

\section{Description of the method}
\label{seecr}
Based on the fundamental use of splines in the signal model
given by Eq.~(\ref{seecrsigmodel}) and the fact that the model 
represents signals that are 
effectively, but not only, chirps, we call the method presented here  
``Spline Enabled Effectively-Chirp Regression" (SEECR).

Some of the principal design choices  behind  SEECR 
are motivated by issues encountered in the  
simpler problem of fitting a smooth curve to noisy data. We briefly review these
issues first before presenting a description of SEECR.

A formal approach to the problem of fitting a smooth curve to noisy data  
is to use regularized least-squares with
a roughness penalty~\cite{green1993nonparametric},
 \begin{eqnarray}
 \widehat{s}(t) & = & \argmin_{s(t)}\sum_{i=0}^{N-1} (y_i -s_i)^2 +\lambda\int_{t_0}^{t_{N-1}}\!\!dt \ddot{s}^2(t)\;.
 \label{roughpenalty}
 \end{eqnarray}
 This method is known as {\em smoothing spline}~\cite{wahba1990spline,hardle1990applied} 
 since the solution 
 turns out to be a cubic spline with the sampling times $t_i$, $i = 0,1,\ldots,N$, 
 as the breakpoints. The influence of the roughness penalty on
  the solution $\widehat{s}(t)$ is controlled by  the 
 regulator gain $\lambda$.
 For $\lambda = 0$, the best fit solution simply matches the data itself, while for
 $\lambda \rightarrow \infty$, it approaches a straight line. Between these two extremes lies
 a solution that is useful for drawing meaningful inferences from the data. 
 
 The natural emergence of splines under a smoothness requirement
 is the main motivation behind our modeling the amplitude envelope
 and instantaneous frequency of a chirp as splines.
 However, estimating these components by directly applying 
 the roughness penalty on them 
 appears to be technically difficult. Instead, 
 we take recourse to another  
 smoothness regularization approach that forms the bases of the 
 {\em regression spline}~\cite{racine2014primer} method. In this method, 
 regularization is achieved by choosing $s(t)$ to be a 
 spline ab initio but limiting the number of 
 breakpoints to be $\ll N$.

 A disadvantage of the regression spline method is 
 that the regularization
 parameter, namely the number of breakpoints, 
 is now discrete and, hence, does not allow fine-grained 
 control over smoothness.
 Moreover, the placement of the breakpoints
 now plays an important role in determining 
 the quality of the fit.
 
 For a predetermined placement of a limited number of breakpoints, the {\em penalized spline} method~\cite{ruppert2003semiparametric}  allows continuous control of smoothness. 
 In the context of the simple curve fitting problem, the penalized spline 
 method models the curve as a linear combination of  B-splines and
 solves
 \begin{eqnarray}
 \widehat{\alpha} & = & \argmin_{\overline{\alpha}} \sum_{i=0}^{N-1} \left(y_i -\overline{\alpha}\mat{A}(\overline{\tau}_a) \right)^2 +\lambda\overline{\alpha}\,\overline{\alpha}^T\;,
 \label{penalizedspline}
 \end{eqnarray}
 where $A_{jm}  =  \mathcal{B}_{j,k}(t_m;\overline{\tau}_a)\;$.

 Finding the optimum placement of breakpoints
 is a challenging non-linear and non-convex
 problem. 
 Methods  proposed in the literature to address this problem generally follow
 the approach of knot insertion and deletion. Only recently have
 optimization methods been developed that are capable of treating breakpoints as 
 completely free parameters. In particular, Particle Swarm Optimization (PSO)~\cite{eberhart1995new,engelbrecht2005fundamentals} has been applied to
 this problem~\cite{galvez2011efficient,mohanty2012particle} 
 and found to have a good performance. 
 
 Along with the placement of breakpoints,
 the number of breakpoints and
  the regulator gain have a significant effect on the quality of estimation. 
For determining the regulator gain,
 Generalized Cross-Validation (GCV)~\cite{golub1979generalized} provides a
 fast method. The number of breakpoints can be selected using the 
 Akaike Information Criterion (AIC)~\cite{Akaike1998}.

SEECR combines the different elements outlined above, namely, penalized 
spline,  GCV, breakpoint optimization using PSO, and AIC. The description of
the algorithm now follows.

\subsection{Regression using Penalized Spline}
\label{regressionmodel}

The signal model in Eq.~(\ref{seecrsigvec}) is estimated
in SEECR by minimizing the penalized least-squares function, 
\begin{eqnarray}
\Lambda(\overline{\alpha},\overline{\theta},\phi_0|\overline{y},\lambda) & = &
R(\overline{\alpha},\overline{\theta},\phi_0|\overline{y}) + \lambda   \overline{\alpha}\,\overline{\alpha}^T\;,
\label{penalizedLS}
\end{eqnarray}
where
\begin{eqnarray}
R(\overline{\alpha},\overline{\theta},\phi_0|\overline{y})  & = &\|
\overline{y} - \overline{s}(\overline{\alpha},\overline{\theta},\phi_0)\|^2\;,\label{rss}
\end{eqnarray}
is the residual norm squared,  over all the signal parameters. 
Henceforth, we drop the explicit listing of parameters wherever it aids clarity.

The positivity of the amplitude envelope, $a(t)\geq 0$,  and B-splines, $\mathcal{B}_{j,k}(t;\overline{\tau}_a) \geq 0$ , $\forall t$, (see Appendix~\ref{app_bsplines}) requires that the minimization of 
$\Lambda$ be performed under a positivity constraint
on $\overline{\alpha}$. 

The estimate of the signal model is obtained using 
the following program of nested minimizations, 
\begin{eqnarray}
\min_{\overline{\alpha},\overline{\theta},\phi_0}
\Lambda & = & \min_{\overline{\theta}}\left(\min_{\phi_0}\left(\min_{\overline{\alpha}} \Lambda\right)\right)\;,
\label{nestedmin}\\
\alpha_i & \geq & 0\;, \forall i\;.
\end{eqnarray}
The order of minimization above, from inner to outer, corresponds to parameters that can be treated semi-analytically to those that need a fully numerical approach.
The steps in solving the program are described below,
starting from the innermost minimization.
As mentioned earlier, the 
regulator gain, $\lambda$, is determined using GCV, which is merged into the 
minimization program at the second step.

\subsection{Innermost minimization}
\label{innermin}

First, we address the unconstrained 
 minimization  over $\overline{\alpha}$. To do so, we  use Eqs.~(\ref{seecrsigvec}) -- (\ref{xmat})
to rewrite $\Lambda$ in a 
more convenient form.
\begin{eqnarray}
\Lambda & = & \|\overline{y}\|^2 +\overline{\alpha}\mat{K}\overline{\alpha}^T 
-2\overline{q}\,\overline{\alpha}^T\;,\\
\mat{K} & = &  
\mat{\Phi}_0 \mat{G} \mat{\Phi}_0^T
\;,
\nonumber\\
{\bf G}& = & {\bf X}{\bf C}^{-1}{\bf X}^T + \lambda \mat{I}\label{gMat}\;,\\
\overline{q} & = & \overline{\eta}\mat{\Phi}_0^T\;,\\
\overline{\eta}  & = &  \overline{y}{\bf C}^{-1} {\bf X}^T 
\;,
\end{eqnarray}
$\mat{K}$ is symmetric and positive definite since  $\overline{x}\mat{K}\overline{x}^T = 
\overline{x}^\prime \mat{G} {\overline{x}^\prime}^T > 0$ for any $\overline{x}$, where
${\overline{x}^\prime}^T = \mat{\Phi}_0 \overline{x}^T$.
It then follows that 
\begin{eqnarray}
\overline{r} & = & \argmin_{\overline{\alpha}}\Lambda  =  \overline{q}\mat{K}^{-1}\;,\label{unconstrainedMinimizer}
\end{eqnarray}
is the solution to the unconstrained inner minimization.

The solution to the constrained minimization problem can be obtained 
from the Karush-Kuhn-Tucker conditions~\cite{kuhn1951}.
These conditions essentially state that the 
solution is either already in the convex cone of $\mathbb{R}^M$
defined by $\alpha_i > 0$, $\forall i$,
or on one of its faces. 

Thus, given the unconstrained minimizer $\overline{r}$ in Eq.~(\ref{unconstrainedMinimizer}),  if $r_i \geq 0$, $\forall i$, then 
$\overline{r}$ itself is the constrained minimizer. 
If not, one has to find the projection 
of $\overline{r}$ on the faces. (The 
inner product to use for the projection is $\langle \overline{x},\overline{w}\rangle =
\overline{x}\mat{K}\overline{w}^T$.)
For this task,  we use the mixed primal-dual 
bases algorithm developed by Fraser and Massam~\cite{fraser1989mixed},
which  returns the edge vectors of the face of the cone that contains the
projection of $\overline{r}$.  

Let the 
projection operator  for the subspace $\mathcal{L}$ spanned by these edge vectors be
$\mat{P}_\mathcal{L}$. Then the solution to the constrained minimization problem is
\begin{eqnarray}
\widehat{\alpha}^T_{\lambda,\phi_0} & = & \mat{P}_{\mathcal{L}} \overline{r}^T\;,
\end{eqnarray}
and the estimated signal at this step in the minimization 
program is,
\begin{eqnarray}
\widehat{s}_{\lambda,\phi_0} & = &  \overline{y}\mat{H}^T_{\lambda,\phi_0}\;,\\
    \mat{H}^T_{\lambda,\phi_0} & = & \mat{C}^{-1}\mat{X}^T\mat{\Phi}_0^T\mat{K}^{-1}\mat{P}_{\mathcal{L}}^T
    \mat{\Phi}_0 \mat{X}\;.
\end{eqnarray}
The subscripts in $\widehat{\alpha}^T_{\lambda,\phi_0}$, $\widehat{s}_{\lambda,\phi_0}$
 and $\mat{H}_{\lambda,\phi_0}$
 make the dependence of these quantities on $\lambda$ and $\phi_0$ explicit. 
 
\subsection{Minimization over $\phi_0$ and GCV}
\label{sec:gcv}
Consider the simpler case where  GCV is used to determine 
$\lambda$ before the minimization over $\phi_0$. 
Let $\lambda_{\rm GCV}(\phi_0)$ be the resulting value. Then,
\begin{eqnarray}
\lambda_{\rm GCV}(\phi_0) & = & \argmin_\lambda {\rm GCV}(\lambda;\phi_0)\;,\\ 
{\rm GCV}(\lambda;\phi_0) & = & \frac{R(\widehat{\alpha}^T_{\lambda,\phi_0},\overline{\theta},\phi_0|\overline{y}) }{
\left(1 - {\rm Tr}(\mat{H}_{\lambda,\phi_0})/N\right)^2}\;,
\label{pregcv}
\end{eqnarray}
where ${\rm Tr}(\mat{H}_{\lambda,\phi_0})$ is the trace of $\mat{H}_{\lambda,\phi_0}$.

Numerical experiments show that computing $\lambda_{\rm GCV}(\phi_0)$ before minimizing $\phi_0$ gives very unstable results. This is because 
the positivity constraint can introduce abrupt changes in the projection 
$\widehat{\alpha}^T_{\lambda,\phi_0}$, by making it switch from
one face of the convex cone to another, as $\phi_0$ 
is varied. 
Independently of this empirical 
reason, 
putting GCV outside the minimization over $\phi_0$ 
also makes sense because it is an approximation to cross-validation, and the right place for the latter is always after minimization over
all relevant signal parameters.

Thus, the  regulator gain in SEECR is determined as follows. 
\begin{eqnarray}
\lambda_{\rm GCV} & = & \argmin_\lambda {\rm GCV}(\lambda;\phi_0(\lambda))\;,
\label{gcv}\\
\phi_0(\lambda) & = & \argmin_{\phi_0}\Lambda(\widehat{\alpha}_{\lambda,\phi_0},\overline{\theta},\phi_0|\overline{y})\;.
\label{minphi0}
\end{eqnarray}
Both of the minimizations above are performed numerically.

\subsection{Outer Minimization}
\label{outermin}

Let
\begin{eqnarray}
\widehat{\alpha} & = & \widehat{\alpha}_{\lambda_{\rm GCV},\phi_0(\lambda_{\rm GCV})}\;,
\end{eqnarray}
 and let the corresponding value of $\Lambda$ be denoted by 
\begin{eqnarray}
F(\overline{\theta}|\overline{y}) & = & \Lambda(\widehat{\alpha},\overline{\theta},\phi_0(\lambda_{\rm GCV})|\overline{y})\;,
\label{fitfunc}
\end{eqnarray}
which we call the {\em fitness function} in the following.
The 
next step in the program given by Eq.~(\ref{nestedmin}) is the minimization of 
the fitness function over the 
 parameters $\overline{\tau}_a$, $\overline{\nu}$,and $\overline{\tau}_f$. 
 
There are two principal challenges in this task. One is the high dimensionality,
given by $M + 2K$, of
the search space, and the other is the degeneracy caused by different permutations of 
the breakpoint sequences giving rise to the same splines. Degeneracies create strong 
local
minima which increase the difficulty of locating the global minimum.

To address the issue of high dimensionality, we lower the number of parameters as follows.
First, we set $[\overline{\tau}_f ]_0 = \tau_{a,0}$ and $[\overline{\tau}_f ]_{K-1} = \tau_{a,M-1}$ because the amplitude
envelope spline, hence the signal itself,
is zero outside the interval $\left[ \tau_{a,0},\tau_{a,M-1}\right]$. 

Secondly, based on the 
Cramer-Rao lower bound on the estimation error in the amplitude  of a monochromatic signal being higher than its frequency,
we can expect that the error in the estimation of the amplitude envelope $a(t)$ of
a chirp is higher than its instantaneous frequency $f(t)$. (This is illustrated later
in Sec.~\ref{seecr_detestperf_150914}.)
A corollary is that one need not invest as much effort in 
modeling the $a(t)$ spline as the 
$f(t)$ one. 
Therefore, we can simplify the 
placement of breakpoints for $a(t)$ considerably, and 
we do so by spacing them uniformly. 
This reduces the number of free $a(t)$
breakpoints from $M$ to 
just two, namely, $\tau_{a,0}$ and $\tau_{a,M-1}$. 
The 
total dimensionality of the search space for the outer minimization
now reduces to $2K$: The 
two end breakpoints for the $a(t)$ spline, the $K-2$ interior breakpoints for the $f(t)$ spline, and the $K$ instantaneous frequency values in $\overline{\nu}$. 

 One
approach to addressing the
issue of degeneracy arising from the permutation symmetry of breakpoints is
to constrain the two breakpoint sequences to be monotonic. That is,
enforce $\tau_{a,M-1} > \tau_{a,0}$
and $[\overline{\tau}_f]_i > [\overline{\tau}_f]_j$ for $i > j$ when searching for the minimum of the 
fitness function. However, this means that the search volume 
no longer has the simple shape of a box, a factor that  
 is known to be detrimental to the performance of PSO. 

An alternative
is to reparametrize breakpoints such that every point in the new 
search space is guaranteed to be a monotonic sequence. 
For any breakpoint sequence $\overline{\tau} = (\tau_0,\tau_1,\ldots,\tau_{P-1})$,
a simple 
reparametrization that leads to monotonicity is,
\begin{eqnarray}
x_0 & = & \tau_0\;,\\
x_{0<i\leq P-1} & = & \frac{\tau_i -\tau_{i-1}}{t_1 - \tau_{i-1}}\;,
\label{naivescheme}
\end{eqnarray}
The new parameters $\overline{x}=(x_0,x_1,\ldots,x_{P-1})$ for $i>0$ are simply distance ratios, with $x_{i>0}\in [0,1]$ (and $x_0 \in [t_0,t_{N-1})$). The search
space in $\overline{x}$ is a box and no additional constraints are needed to ensure the monotonicity of a breakpoint sequence. 

There is, however, a disadvantage to the reparametrization scheme presented above,
which is that a
uniformly spaced breakpoint sequence is pushed towards
the boundary of the 
box. This is not of much concern for the amplitude envelope spline since
we have reduced the number of free breakpoints to just two. However, 
the variant of PSO used in this paper
is generally known to perform better if a global minimum is located 
towards the central region of a search space. Hence,  its performance  
would suffer with the above reparametrization if an instantaneous frequency 
spline were best represented by uniformly spaced breakpoints.

A clever scheme that circumvents this problem,
while still preserving monotonicity, was proposed in~\cite{calvin-siuro}.
\begin{eqnarray}
x_0 & = & \tau_0\;,\label{clscheme_start}\\
x_{1\leq i \leq P-2} & = & \frac{\tau_i - \tau_{i-1}}{\tau_{i+1}-\tau_{i-1}}\;.
\label{clscheme}\\
x_{P-1} & = & \tau_{P-1}\;,\label{clscheme_end}
\end{eqnarray}
Here, the distance ratios in Eq.~(\ref{clscheme}) are relative to the gap between the 
enclosing knots rather than, as  in Eq.~(\ref{naivescheme}), a knot and the 
end point of the data. 

With the reparametrization in Eqs.~(\ref{clscheme_start}) -- (\ref{clscheme_end}), no obvious degeneracy is left in the
fitness function. However, that does not mean that there are no local 
minima in the fitness function. In fact, as with the estimation of any oscillatory signal,
multiple local minima may be expected that may be scattered widely in the search 
space. Therefore, the search for the global minimum  cannot be performed with deterministic local minimizers and a method such as PSO must be used.
(Despite the reduction in the number of parameters, the dimensionality of
the search space is high enough that grid-based search strategies would simply be 
computationally infeasible.)


\subsection{Model Selection}
\label{AIC}
All of the preceding description relates to fixed numbers, $M$ and $K$ respectively, of breakpoints 
for the amplitude envelope and instantaneous frequency splines. 
The final step in SEECR is an automated determination of their best values using AIC. 
The general expression for AIC  is
\begin{eqnarray}
{\rm AIC} & = & 2 N_{\rm params} - 2 \ln \widehat{L}\;,
\label{aic_original}
\end{eqnarray}
 where $N_{\rm params}$ is the total number of free parameters involved in a given model
 and $\widehat{L}$ is the maximum value, over the space of these parameters,
 of the likelihood function. 
 The best among a set of models is the one that has the minimum AIC value.

 In our case,  $N_{\rm params} = M + 2K +1$, where $M$ is the number
  of B-spline coefficients $\overline{\alpha}$, $2K$ is the total number
  of breakpoints and corresponding instantaneous frequency values 
  (Sec.~\ref{outermin}), 
  and  1 is for the $\phi_0$ parameter. 
  
  For Gaussian stationary noise, the log-likelihood can be expressed as $-2 R(\overline{\alpha},\overline{\theta},\phi_0|\overline{y})$ [see Eq.~(\ref{rss})].
  Hence, maximizing the former
 is equivalent to minimizing the latter.
  In the case of SEECR, $R(\overline{\alpha},\overline{\theta},\phi_0|\overline{y})$ is replaced by $\Lambda(\overline{\alpha},\overline{\theta},\phi_0|\overline{y})$ [see Eq.~(\ref{penalizedLS})]. Its minimization over the parameters $\overline{\alpha}$ and $\phi_0$
  yields the fitness function, $F(\overline{\theta}|\overline{y})$, defined in Eq.~(\ref{fitfunc}).
 Thus, $-2\ln\widehat{L}$ in Eq.~(\ref{aic_original}) is replaced by the minimum value, $\widehat{F}_{M,K}$, of
  the fitness function,  
 \begin{eqnarray}
 \widehat{F}_{M,K} & = & \min_{\overline{\theta}} F(\overline{\theta}|\overline{y})\;.
 \label{bestfitval}
 \end{eqnarray}
  Hence, the value of AIC in 
 our case is 
 given by
 \begin{eqnarray}
 {\rm AIC} & = & 2(M+2K) + \widehat{F}_{M,K}\;,
 \label{aic}
 \end{eqnarray}
 where we have dropped constants that do not affect the minimization of AIC. The number of breakpoints in the model that minimizes ${\rm AIC}$ 
 will be denoted by $\widehat{M}$ and $\widehat{K}$ in the following.
\subsection{Amplitude Envelope and Instantaneous Frequency Estimates}
\label{sec:estimQuant}
Let the final estimated signal sequence, obtained 
from the best model selected by AIC, be denoted by 
$\widehat{s}$. To obtain the best fit sequences for the amplitude envelope, $\widehat{a}$,
and instantaneous frequency, $\widehat{f}$, we construct the analytic sequence $\widehat{s}^{\rm (anlt)}$,
\begin{eqnarray}
\widehat{s}^{\rm (anlt)} & = & \widehat{s} + i \mathcal{H}[\widehat{s}]\;,
\end{eqnarray}
where $\mathcal{H}$ is the discrete Hilbert transform~\cite{oppenheim_Schafer_dsp} operator. 
Then 
\begin{eqnarray}
\widehat{a}_j & = & |\widehat{s}^{\rm (anlt)}_j|\;,\\
\widehat{f}_i & = & \frac{\widehat{\phi}_{i+1}-\widehat{\phi}_i}{t_{i+1}-t_i}\;,\\
\widehat{\phi}_j & = & {\rm arg} (\widehat{s}^{\rm (anlt)}_j)\;,
\end{eqnarray}
where $j=0,1,\ldots,N-1$, $i = 0,1,\ldots,N-2$, and  continuity is enforced 
across jumps of $\pm \pi$ in $\widehat{\phi}$.

We do not obtain $\widehat{a}$ and $\widehat{f}$ directly from their respective estimated splines because the two 
 interact non-linearly
 in $\widehat{s}$ to give a better estimate of the signal than what is possible with the 
 splines alone. 
 However, a minor downside of using the Hilbert transform is that it creates artifacts in 
 $\widehat{f}$. Usually these are samples that are negative or very close to the 
 Nyquist rate, and easily eliminated by setting them to zero. The 
 $\widehat{a}$ sequence generally does not present such artifacts.
 
 In the following, exactly the same process as above is used to get the amplitude envelope and instantaneous frequency of the true signal.

\subsection{Evolution of SEECR}
As mentioned earlier, SEECR is the culmination of a sequence of intermediated 
methods~\cite{mohanty2012particle,2012AstRv...7d...4M,mohanty2014}. Here, we 
briefly summarize the similarities and differences between SEECR and the 
preceding methods.

In~\cite{mohanty2012particle}, the simple problem of fitting data with a spline was considered. Thus, the signal model used was,
$s(t;\overline{\alpha},\overline{\tau}_a)  =  a(t;\overline{\alpha},\overline{\tau}_a)$,
with $a(t;\overline{\alpha},\overline{\tau}_a)$ given by Eq.~(\ref{aoft_spline}).
PSO was proposed for optimizing the residual norm squared [Eq.~(\ref{rss})] over $\overline{\tau}_a$ without a monotonicity ($[\overline{\tau}_a]_{i>j} > [\overline{\tau}_a]_j$) constraint. (In addition, \cite{mohanty2012particle} uses a  variant
of PSO that is different from the one used in SEECR.)

The signal model used here [Eq.~(\ref{seecrsigmodel})] was introduced in a more restricted form in~\cite{2012AstRv...7d...4M}: it was assumed that $\phi(t)\rightarrow 2\pi f_0 t + \phi(t)$, with $\phi(t)$ changing over a much longer timescale than the period, $1/f_0$, of the carrier. This restriction allows the signal to be heterodyned, yielding the two
quadratures $a(t)\cos\phi(t)$ and $a(t)\sin\phi(t)$. The method in~\cite{mohanty2012particle} was then used to estimate the quadratures independently. 
While  the importance of modeling both the amplitude and phase evolution of a signal through splines was emphasized in~\cite{2012AstRv...7d...4M}, the heterodyning approach 
is completely different from what is done in SEECR.

In~\cite{mohanty2014}, the signal model was generalized to essentially match  Eq.~(\ref{seecrsigmodel}). However, the initial phase parameter, $\phi_0$, was not 
included in the model, which simplifies the steps involved in
Sec.~\ref{innermin} considerably.
In addition $f(t)$ was modeled with a linear, not cubic, spline. The
number of breakpoints, $M$ and $K$, were not varied and model selection (see Sec.~\ref{AIC}) was not used. The use of GCV was introduced but did not face the 
complication, described in Sec.~\ref{sec:gcv}, involved in meshing it 
with the minimization over $\phi_0$.

 
 \section{Description of the simulations}
 \label{sims}
 We quantify the performance of SEECR using statistically independent simulated data
 realizations corresponding to the data model in Eq.~(\ref{datamodel}). 
 $H_0$ data realizations are drawn from a zero mean Gaussian
 white noise process with unit variance (i.e., an i.i.d $N(0,1)$ sequence).
 There is no loss of generality because the inner product in Eq.~(\ref{innprod}) is
 equivalent in the Fourier domain 
 to the Euclidean inner product of a white noise sequence with a whitened 
 signal. Since the choice of waveforms 
 for unmodeled signals is arbitrary to begin with, they can
 be assumed to be those of the whitened signals.
 
 \subsection{Simulated Signal Waveforms}
 \label{simwaveforms}
 
 We use the following simulated signal waveforms, covering a wide range in the behavior of 
 the amplitude envelope and instantaneous frequency.
 Each signal is assigned a label followed by pertinent information about it. 
 For the signals where expressions for $a(t)$ and $\phi(t)$ are given, $s(t) = a(t) \sin(\phi(t))$. 
 We have taken care to set some of the signal parameters, such as the start time or the carrier frequency, at values that are not related in a special way to the sampling grid in either the temporal or the Fourier domain.
 All data realizations containing the signals listed below have a duration of $2.0$~sec
 with a sampling frequency of $4096$~Hz.
  \begin{description}
 \item{TS}: Transient sinusoid with $a(t)=1$ for $t\in [0.4,1.4]$~sec and zero otherwise.  $\phi(t) =2\pi f_0 (t-t_0)$, with $f(t)=f_0=473.0$~Hz, and $t_0 = 0.4$~sec.
 \item{SG}: Sine-Gaussian signal with constant
 $f(t) =f_0= 204.8$~Hz and $a(t)$ having a Gaussian shape
 that is symmetric with respect to the mid-point  of the signal. 
  $a(t) = \exp\left(-(t-t_0)^2/(2\times ({\rm FWHM}/2.355)^2)\right)$, for $t\in [0.4,0.9]$~sec and zero otherwise. The peak of $a(t)$ is at $t_0 = 0.65$~sec 
 and ${\rm FWHM}=0.29$~sec is its full width at half maximum. $\phi(t) = 2\pi f_0 (t-t_0)+\pi/2$.
 \item{3PS}: Monochromatic signal with three Gaussian peaks in the amplitude envelope.
 This signal is obtained by  concatenating three SG signals. (The middle signal is the negative of the SG in order to reduce the effect of phase discontinuities at its boundaries.) $a(t) \neq 0$ for $t\in [0.3,1.8]$~sec and zero otherwise.
 \item{LC}: Linear chirp (quadratic phase) with constant amplitude. 
 $\phi(t) = 2\pi (f_0 t + f_1 t^2)$, with $f_0 = 200$~Hz and $f_1=300$~Hz$^2$, and
 $a(t) = 1$ for $t\in [0.4,1.4]$~sec and zero otherwise. 
\item{QC}:  Quadratic chirp (cubic phase) with constant amplitude as defined in~\cite{candes2008gravitational}.
$\phi(t) = 2\pi (f_0 t+f_1 t^3$), where $f_0 = (2\pi)^{-1} 256 $~Hz and $f_1 = (2\pi)^{-1}(512/3)$~Hz$^3$. $a(t) = 1$ for $t\in [0.4,1.4]$ and zero otherwise.
The start and end frequencies are $40.7$~Hz and $122.2$~Hz respectively.
\item{CC}: Cosine phase chirp with cosine modulated amplitude as defined in~\cite{candes2008gravitational}. 
$a(t)=2+\cos(2\pi f_0 (t-t_0)+\pi/4)$ for $t\in [0.4,1.4]$~sec and zero otherwise. 
Here, $f_0 = 1$~Hz and $t_0 = 0.4$~sec. $\phi(t) = \phi_m\sin(2\pi f_0 (t-t_0))
+ 2\pi f_1 (t-t_0)$, with $\phi_m = 1024/\pi$~rad and $f_1 = 400$~Hz.
The behavior of $f(t)$ can be seen from its spectrogram in Fig.~\ref{CC_sig_specg}.
(See Appendix.~\ref{app:spectrogram} for the precise definition of a spectrogram as 
used in this paper.)
 \item{s11WW}: A CCSN waveform obtained from~\cite{stellarcollapse_dot_org} corresponding to the accoustic supernova model~\cite{2006PhRvL..96t1102O}. The waveform time series was anti-aliased and downsampled  to $f_s = 4096$~Hz, leaving no discernible changes
 as most of the power in the signal
 lies below $\sim 1.5$~kHz. Both $a(t)$ and $f(t)$ have a complex evolution for
 this waveform due to the simultaneous presence of multiple chirping components as can be seen from the spectrogram of this signal
 in Fig.~\ref{s11WW_sig_specg}.
 However, there is 
 a single component that dominates in power, making the single chirp model assumed
 in SEECR a good fit. In each data realization, the signal starts at $t=0.4$~sec and
 terminates at $1.173$~sec.
  \end{description}
 \begin{figure}
     \centering
     \includegraphics[scale=0.45]{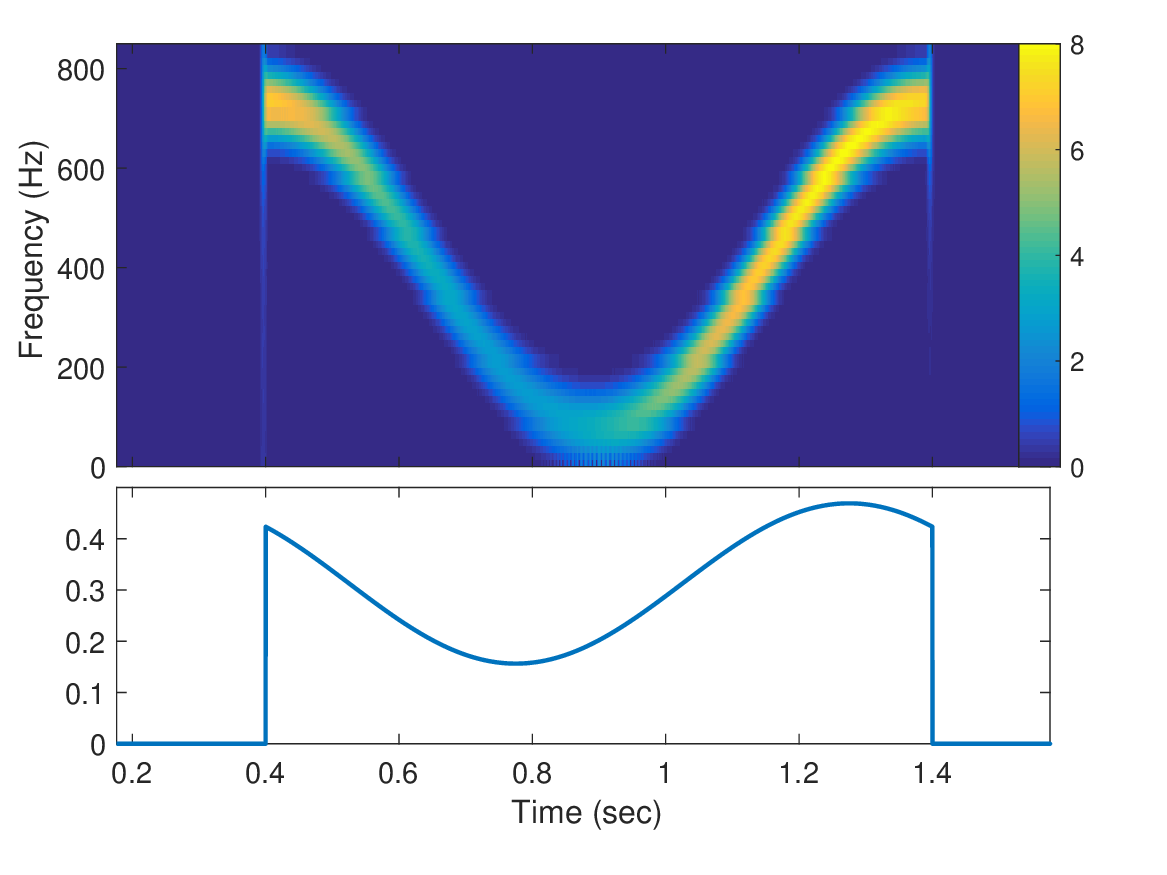}
     \caption{(Top) Spectrogram of the CC signal showing the behavior of its 
     instantaneous frequency $f(t)$.  (Bottom) The amplitude envelope, $a(t)$, of the CC signal 
     for ${\rm SNR} =15$. The locations of the minima in $f(t)$ and $a(t)$ do not coincide, 
     leading to the lopsided distribution of signal power, with the part after the minimum in $f(t)$ being stronger.}
     \label{CC_sig_specg}
 \end{figure}
\begin{figure}
     \centering
     \includegraphics[scale=0.3]{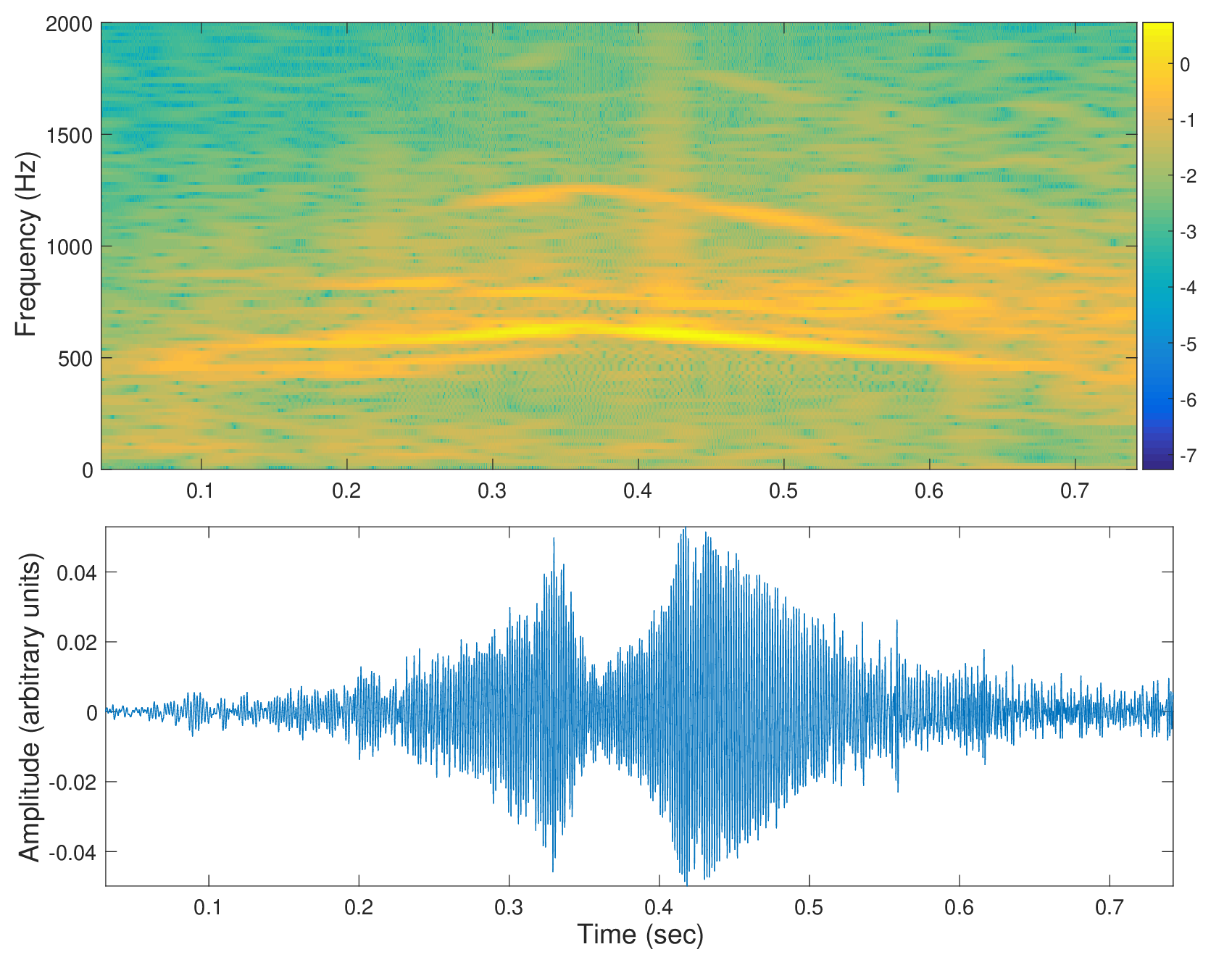}
     \caption{(Top) Spectrogram of the s11WW signal with the
     magnitude shown on a log-scale
 in order to elucidate the multiple chirping components more clearly. (Bottom) The signal time
 series where the amplitude has been scaled such that ${\rm SNR} = 1$.}
     \label{s11WW_sig_specg}
 \end{figure}
 
 When constructing an $H_1$ data realization, the signal amplitude is 
normalized such that it has a certain matched filtering 
 signal to noise ratio (SNR). The SNR of a signal 
 characterizes the performance of the 
 optimal statistic, namely the log-likelihood ratio (LLR), for the binary 
 hypotheses test where there is only one signal waveform and it is completely known
 {\em a priori}. For the Gaussian white noise process used in the simulations, 
 \begin{eqnarray}
 {\rm SNR} & = & {\frac{E[{\rm LLR} | H_1]- E[{\rm LLR}|H_0]}{
    \left[E[({\rm LLR}-E[{\rm LLR}|H_0])^2|H_0]\right]^{1/2}}}\;,\nonumber\\
    & = &  \left[ \sum_{i = 0}^{N-1} s_i^2 \right]^{1/2}\;,
 \end{eqnarray}
 where $E[{\rm LLR}|H_{i}]$, $i = 0$ or $1$,  denotes  expectation 
 under hypothesis $H_i$.
 For generating data realizations under $H_1$, we use three SNR values,
  ${\rm SNR}=10, 12, 15$, for each of the simulated signals.
 
 \subsection{GW150914 Analysis}
 \label{realsignals}
 The simulated waveforms listed so far have durations of $\geq 1$~sec, with the exception of SG that has a
 duration of $0.5$~sec. Although the main target for SEECR are signals in this duration range, it is 
 interesting to quantify its performance for a significantly shorter chirp. 
 
 For this purpose, we simply 
 use the real event, GW150914, which furnishes a chirp of duration $< 0.2$~sec. 
 However, GW150914  
 had an exceptionally high 
 observed network SNR of $24$, and a single detector SNR of $\approx 20$~\cite{PhysRevD.93.122003} in the
 Hanford detector, making it an easy case for burst search algorithms.
 To test if SEECR could have detected this
 signal at weaker strengths,
 we use the real  GW150914 data as a seed to generate new realizations in which the
 observed SNR is reduced to $\approx 10$. 
 
\begin{figure}
     \centering
     \includegraphics[scale=0.5]{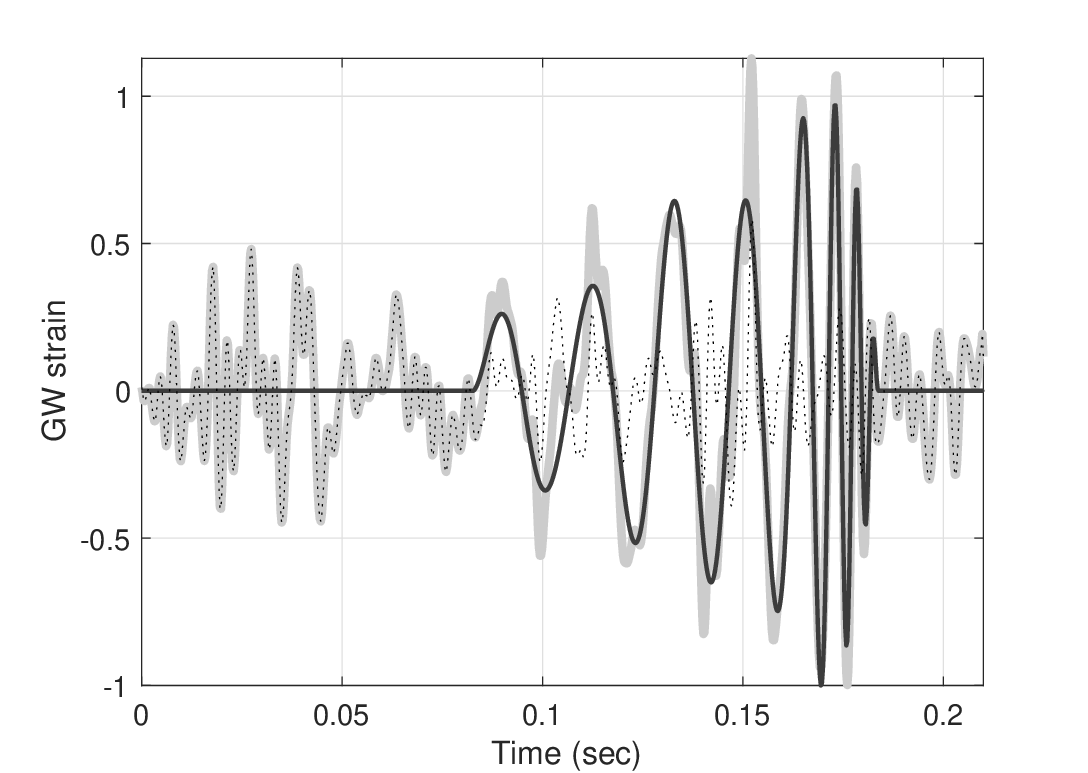}
     \caption{The thick gray curve is the 
     GW150914 data from the Hanford detector. The solid black curve is 
     the signal estimated by SEECR. The dashed curve shows the residual after
     subtracting the estimated signal from the data.}
     \label{fig:gw150914_seecr_est}
 \end{figure}

 First, we  take 
 the time series from the Hanford detector, shown in Fig.~1 of~\cite{PhysRevLett.116.061102}
 and estimate the standard deviation 
 of the noise in the data. This is done by
 estimating the signal using SEECR 
 and subtracting it from the data to obtain the residual. Fig.~\ref{fig:gw150914_seecr_est} shows the data, the
 estimated signal, and the residual.  
 The residual has a sample standard deviation of 
 $0.16$.
 
 Next, 
 a realization of pseudo-random 
 noise is generated and added to the 
  original data. 
  The noise realization  is first generated as white noise with unit variance and then low-pass filtered, using
 an order $40$ Finite Impulse Response filter, to the band $[0,450]$~Hz. The resulting time series, having a standard deviation of $\sigma_{\rm filt}$, is then scaled by $[\sqrt{3}\times 0.16/\sigma_{\rm filt}]$. 
 Modulo the sampling error in the standard deviation estimate,  the observed ${\rm SNR}$ of the signal in the new
 realization is reduced by a factor of 2.
 To generate $H_0$ data, we follow the same procedure but use a 
 scaling factor of  $2\times 0.16/\sigma_{\rm filt}$.
 
\begin{figure}
     \centering
     \includegraphics[scale=0.43]{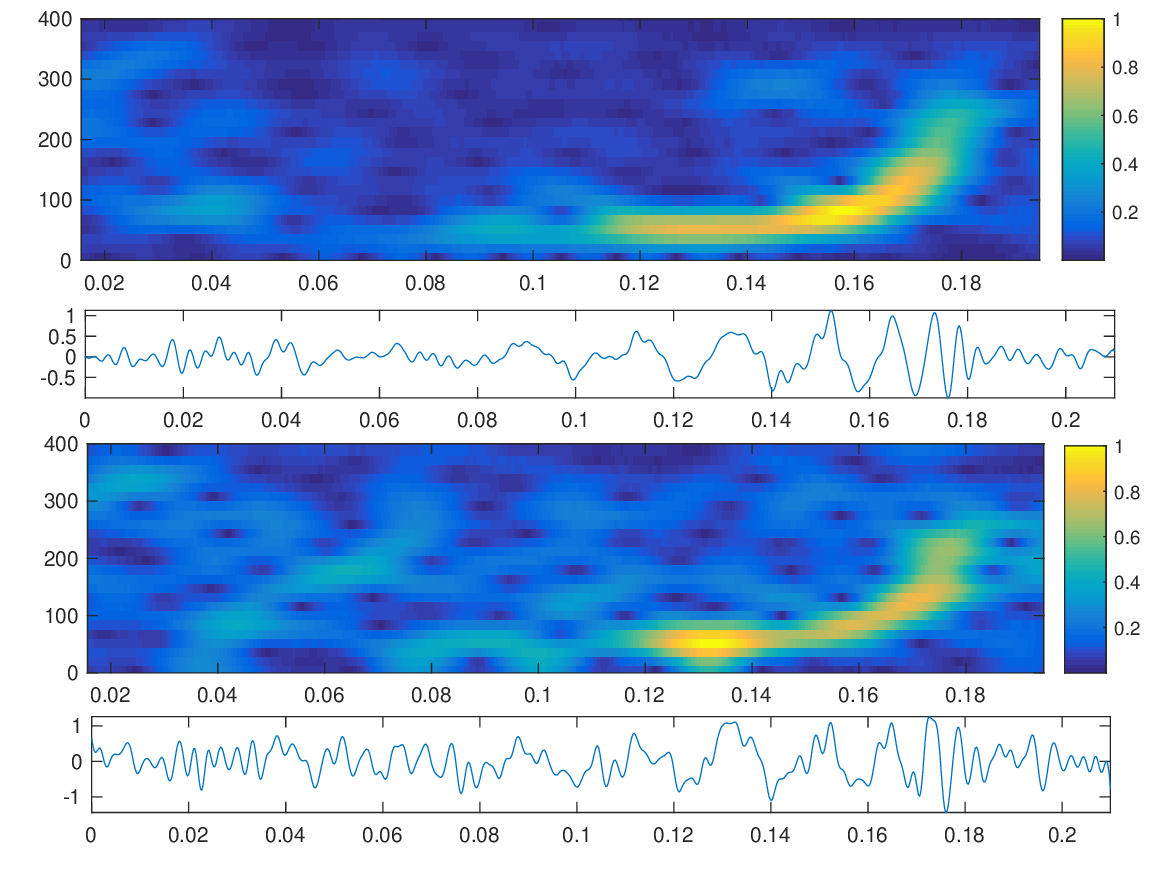}
     \caption{GW150914 data from the Hanford detector compared with a realization obtained by adding excess pseudo-random white noise. The first and second panels from the top
     show the spectrogram, obtained with a window length of 128 and overlap between consecutive windows of 127 samples, and the data time series respectively. The bottom two panels are the corresponding plots for a data realization where the observed SNR has been reduced by a factor of 2. In all panels, the horizontal axis shows time in seconds. The vertical axes in the case of spectrograms shows frequency (Hz). In the time series plots, the vertical axis shows the (whitened) GW strain ($\times 10^{-21}$).}
     \label{gw150914_spgrm_tser}
 \end{figure}

 As with the simulated
 signals, independent
 realizations of $H_1$ and $H_0$ data are generated for GW150914.
 Each data realization has a duration of $0.21$~sec with a sampling frequency of $4096$Hz. Fig.~\ref{gw150914_spgrm_tser} 
 compares the original data with one such 
 realization.

 \subsection{SEECR Parameter Settings}
 \label{psoparams}

The principal user-determined parameters governing
SEECR are the number of breakpoints, $M$ and $K$, for 
 the amplitude envelope
and instantaneous frequency splines
respectively. The user provides a set of values for $M$ and $K$ and, as described in Sec.~\ref{AIC}, AIC is
used to pick the best combination. 

In principle,  one need only specify
the maximum values of $M$ and $K$ and let AIC examine all the integers below them. 
However, this  is wasteful 
since the  signal 
estimates, hence the AIC values,
may not differ much between nearby models. 
This is particularly true at higher values for the number of 
breakpoints where
nearby models start differing less and less in their fit quality. 
Hence, computational costs 
can be reduced substantially by spacing models out judiciously.

Based on the above and keeping computational costs in mind, we arrived at the sets
$\{5, 6, 7, 9, 11\}$ and $\{3, 4, 5, 7\}$
for $M$ and $K$ respectively, resulting in 
$20$ different models, that are kept fixed throughout this paper. 

Besides the above parameters, 
there are the parameters associated with PSO and
the range, $[\nu_{\rm min},\nu_{\rm max}]$, for the $K$
instantaneous frequency values $\nu_i$, that it needs to search.
(The range for the amplitude envelope end breakpoint parameters is set so that the 
entire data segment is covered.)
A virtue of the PSO algorithm is the robustness of its parameter
 settings. This
 allows us to simply keep the same settings~\cite{bratton2007defining} as used in~\cite{2015ApJ...815..125W}, to which we refer the reader for further details.
 For the above set of $K$ values, the dimensionality of the search space for PSO ranges between 6 and 14. 

Like
all stochastic global optimizers, PSO is not guaranteed to converge to the global 
minimum. However, the probability of success can be increased exponentially
by doing multiple runs of PSO, with statistically independent
initial states, on the same data realization
and picking the run that returns the best
fitness value.  
The number of independent PSO runs is set to 8 in this paper.

We keep $\nu_{\rm max}$ slightly below 
 the Nyquist frequency of 
the data to prevent too many PSO particles from escaping the search region to explore 
 physically invalid frequencies.
 Except Sec.~\ref{seecr_cpp_trksrch_detestperf}, where $\nu_{\rm max} = 510$~Hz,
 we set $\nu_{\rm min}=0$ and 
 $\nu_{\rm max} = 2000$~Hz.  
 
 Finally, we use splines of order 4 (cubic splines) for both the amplitude 
 envelope and instantaneous frequency.

 
 \section{Results}
 \label{results}
 The presentation of the simulation results is organized as follows. In Sec.~\ref{seecr_detperf}, we focus on the
 detection performance of SEECR.
 Sec.~\ref{seecr_estperf} describes its performance in estimating 
 the amplitude envelope and instantaneous frequency of a signal. 
  The results in Sec.~\ref{seecr_detperf}
 and Sec.~\ref{seecr_estperf} use the
 set of signals described in Sec.~\ref{simwaveforms} with 
 $500$ realizations of $H_0$ and a minimum of $50$ realizations of $H_1$ data for each signal and each SNR.
 Sec.~\ref{seecr_detestperf_150914} 
 presents results from 
  the GW150914 analysis described in 
  Sec.~\ref{realsignals}.
  For these results
  we use 100 $H_0$ and 50 $H_1$ data realizations.
 
\subsection{Detection performance}
\label{seecr_detperf}
\begin{figure}
    \centering
     \includegraphics[scale=0.25]{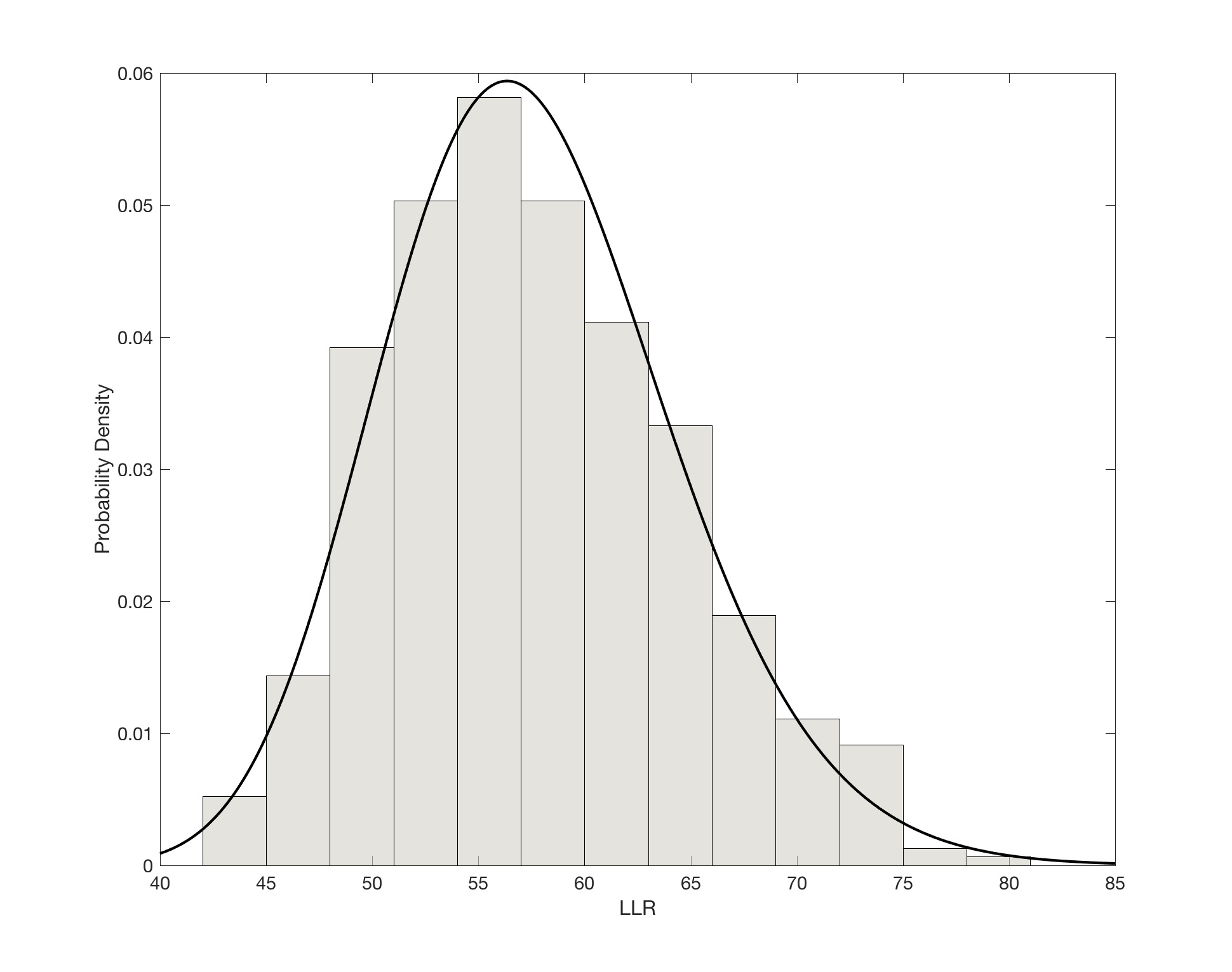}
     \caption{Estimated distribution of the SEECR detection statistic, LLR, under
     the null hypothesis. The distribution is estimated from 500 realizations of an i.i.d $N(0,1)$ sequence with 8192 samples. The
     bars show the histogram, with the count in each bin normalized to represent the
     probability density function (pdf). The solid curve shows the best fit lognormal pdf,
     obtained for $\mu=4.04563$, and $\sigma = 0.11836$.}
     \label{H0DistLLR}
\end{figure}

For SEECR to function as a detector, we must choose a detection
statistic, and a natural choice for it is
the LLR evaluated at the best fit model.
Following the discussion in Sec.~\ref{AIC} regarding the relation between log-likelihood
and $\widehat{F}_{\widehat{M},\widehat{K}}$ [defined in Eq.~(\ref{bestfitval})], 
\begin{eqnarray}
{\rm LLR} & = & \|\overline{y}\|^2 - \widehat{F}_{\widehat{M},\widehat{K}}\;,
\label{eq:detstatistics}
\end{eqnarray}
To obtain the threshold corresponding to
a given false alarm probability, we estimate the
probability density function (pdf) of LLR from the $H_0$ data realizations. 
Fig.~\ref{H0DistLLR} shows the estimated pdf
along with the best fit lognormal pdf. We pick the lognormal pdf,
\begin{eqnarray}
p(x) & = & \frac{1}{x\sigma\sqrt{2\pi}}\exp\left(-\frac{(\ln x -\mu)^2}{2\sigma^2}\right)\;,
\end{eqnarray}
because it provides a good match to the asymmetry of the estimated distribution
around its mode, as well as its heavy tail, with only two free parameters.

We quote detection probabilities at two 
values of the false alarm probability: $1/500 = 2\times 10^{-3}$
and $2\times 10^{-4}$.
Since each data realization is 2~sec long, the values of the 
false alarm rate (FAR) are $10^{-3}$~events/sec and 
$10^{-4}$~events/sec respectively.
(The resulting FAR for coincidence based detection between a pair of GW detectors is discussed in Sec.~\ref{conclusions}.)
The corresponding
thresholds on ${\rm LLR}$ obtained from the best fit lognormal are $80.3$ and $86.9$ respectively.

Table~\ref{detprobtable} reports the detection probabilities  for
the simulated signals in Sec.~\ref{simwaveforms} at the different ${\rm SNR}$ values
used in this study.
The
error interval associated with each detection probability corresponds to $\pm 1\sigma$, where 
\begin{eqnarray}
\sigma & = & [p(1-p)/N_{\rm trials}]^{1/2}\;,
\label{eq:errorbar_prob}
\end{eqnarray}
with $p$ being the estimated detection probability and $N_{\rm trials}$ being the number of $H_1$ data
realizations used. Note that the estimated detection probability does not have a Normal
distribution, and the error interval above 
is not strictly appropriate, for $p$ close to unity or
zero. For extreme values of $p$, one may use the Clopper-Pearson confidence interval~\cite{thulin2014cost} to assess the error in $p$. In our case, the extreme value of concern is $p = 1$, for which the interval is given by $((\alpha/2)^{(1/N_{\rm trials})},1)$, where $\alpha$ is the confidence level. For $\alpha = 0.95$ and $N_{\rm trials} \geq 50$, the interval is $(\geq 0.9852,1)$.
Here, and in the rest of the paper, a quoted estimated detection probability of unity is understood to refer to the above confidence interval.
\begin{table}
\begin{tabular}{|c|c|c|c|c|}\hline
Signal &\multicolumn{2}{c|}{FAR = $10^{-3}$~events/sec}  & 
\multicolumn{2}{c|}{FAR = $10^{-4}$~events/sec}\\
\cline{2-5}
 & SNR=10 & SNR=12 & SNR=10 & SNR=12  \\
\hline
 TS &		0.98 $\pm$ 0.02	 &		1.00 $\pm$ 0.00 &0.96 $\pm$ 0.03 & 1.00 $\pm$ 0.00\\
SG 	 &	1.00 $\pm$ 0.00	 &	1.00 $\pm$ 0.00 & 0.98 $\pm$ 0.02 & 1.00 $\pm$ 0.00\\
3PS  &		0.84 $\pm$ 0.05	 &		0.94 $\pm$ 0.03 & 0.82$\pm$ 0.05 & 0.92 $\pm$ 0.04\\
LC  &		0.52 $\pm$ 0.07	 &		0.90 $\pm$ 0.04 & 0.40 $\pm$ 0.07 & 0.84 $\pm$ 0.05\\
QC	 &  	0.61 $\pm$ 0.04	 &		0.97$\pm$ 0.01 & 0.48 $\pm$ 0.05 & 0.95 $\pm$ 0.02\\
CC	 &	0.22 $\pm$ 0.04	 &  	0.68 $\pm$ 0.04 & 0.092 $\pm$ 0.03 & 0.53 $\pm$ 0.05\\
s11WW	 &	0.86 $\pm$ 0.05 &		0.98 $\pm$ 0.02 & 0.72 $\pm$ 0.06 & 0.98 $\pm$ 0.02\\
\hline
\end{tabular}
\caption{\label{detprobtable}
Estimated detection probabilities, and their $1\sigma$ error intervals at two different  false alarm rates.  
The detection probability 
for each SNR value is estimated using $50$ $H_1$ 
data realizations except for QC and CC, where the number of realizations is $120$.
The detection probability at ${\rm SNR} = 15$ is unity at 
both FARs for every signal and, hence, not listed 
explicitly. }
\end{table}

\begin{figure}
    \centering
     \includegraphics[scale=0.42]{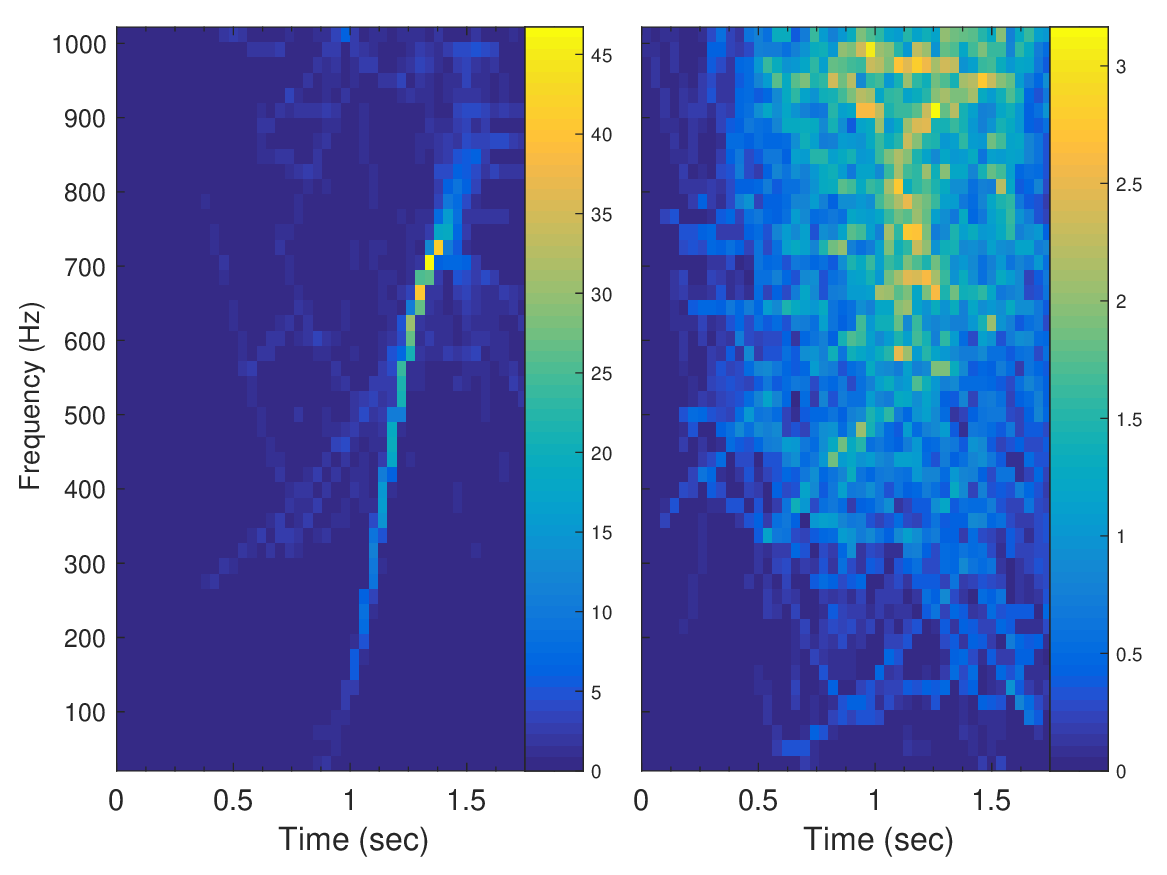}
     \caption{Two dimensional histogram of estimated instantaneous frequencies for (left panel)
     $H_1$ data realizations containing the CC signal with ${\rm SNR}=10$, and (right panel) $H_0$ data. Each histogram 
     is constructed by plotting all the estimated frequencies and counting the number of plotted
     points in a regular grid of 2D bins. There are 50 bins along each dimension. For the left panel, only those data realizations are included that had LLR values less than the detection 
     threshold. All realizations of $H_0$ data are used for the right panel. The counts in each panel are normalized by the respective number of trials used.}
     \label{CC_instf_dist}
\end{figure}

We see that, at ${\rm SNR}=10$
and a FAR of $10^{-3}$~events/sec, SEECR attains a 
detection probability of $\gtrapprox 0.5$ for all the signals
except CC. 
 In itself, the reduced power for this signal 
 is not surprising given that it has the most extreme 
amplitude and instantaneous frequency variation.
However, an additional reason
appears to be the lopsided distribution of signal
power as seen in Fig.~\ref{CC_sig_specg}. Its effect on the estimated signal is shown in Fig.~\ref{CC_instf_dist}. We see that the estimated instantaneous frequency tends to match
only the part of the signal that is louder, and the initial half of the signal is 
missed completely. 

Fig.~\ref{CC_instf_dist} also shows the estimated instantaneous 
frequencies from $H_0$ data. It is interesting that 
the imprint of the signal on the distribution of   
estimated 
instantaneous frequencies is quite clear even at ${\rm SNR}=10$.

The detection probability of the CC signal is reduced 
substantially for the FAR of $10^{-4}$~events/sec but then 
climbs to  $0.53\pm 0.05$ at ${\rm SNR}=12$. SEECR achieves
a detection probability $\gtrapprox 0.8$ for all the other signals at 
${\rm SNR}=12$ for this FAR. 


\subsection{Estimation performance}
\label{seecr_estperf}
Gauging the performance of any method on the estimation of chirp signals requires
metrics that go beyond the simple mean squared error (MSE)
$\| \overline{s} - \widehat{s}\|^2$ between the true signal, $\overline{s}$, and its estimate $\widehat{s}$. This is because, as discussed in Sec.~\ref{outermin}, the error in estimating the amplitude envelope $a(t)$ of a chirp 
can be significantly higher than that for its instantaneous frequency $f(t)$ but
they are conflated in the MSE without any kind of weighting.
 Moreover, going by the case of binary 
inspiral signals, most of the physically important information carried by a GW chirp is likely to reside in $f(t)$ and one would like to study the error in estimating it 
independently of $a(t)$. 
This motivates the introduction of a set of
metrics to separately quantify the estimation 
performance for $a(t)$ and  $f(t)$. 

 \subsubsection{Estimation metrics}
 \label{sec:estimation_metrics}
 The metrics proposed here are based on the physically relevant information
 one would like to extract from any estimated signal. At the most
 basic level, this consists of  
 the time of arrival, the duration, and as much of $f(t)$
 as possible.  
 
In a parametric search method, the time of arrival and duration
are explicit parameters of the signal model and are measured
as such. In the
case of unmodeled chirps, however, the measured quantities are
$a(t)$ and $f(t)$, and the time of arrival and duration
must be derived from 
them. 
Among the two, it is natural
to use $a(t)$ for this inference but 
due account must be taken of estimation error, which can be expected to be higher where the true $a(t)$ is smaller. Therefore, for example, simply using the start time of the estimated $a(t)$ as the time of arrival is not a good idea because the start of a signal is precisely where the true $a(t)$ decays to zero and the estimation 
 error is likely to be highest.
 
Consider a finite duration 
amplitude envelope $a(t)$, with $a(t) = 0$ for $t\notin [t_1, t_2]$. 
Given that $a(t)\geq 0$ everywhere and integrable, 
one can normalize it to construct a pdf over $t$,
\begin{eqnarray}
p_a(t) & = & \frac{a(t)}{\int_{t_1}^{t_2}dt\, a(t)}\;.
\end{eqnarray}
We define the time of arrival, denoted by $t_{a}$, 
 as the median of this pdf,
\begin{eqnarray}
\int_{t_1}^{t_{a}} p_a(t) dt & = & \frac{1}{2} \;.
\label{mutmed}
\end{eqnarray}
The duration $t_{D}$ is defined 
as the 
inter-quartile range (IQR) -- the difference between the first and the third quartiles --
of the pdf,
\begin{eqnarray}
t_{D} & = &q(0.75) - q(0.25)\;,\\
\int_{t_1}^{q(\alpha)}dt\, p_a(t) & = & \alpha\;.
\end{eqnarray}
The median is preferable to the mean of $p_a(t)$ as an estimator of $t_a$
because it is more robust against the increased error in the tails of $p_a(t)$ near the start and end of a signal. Generally, these errors need
not be equal at the two ends, giving rise to a larger bias in the mean than in the median. For the same reason, the IQR is a more robust measure of the duration
than the standard deviation. For reference, the IQR of a normal distribution with
standard deviation $\sigma$ is $1.34\sigma$.

We denote 
the metrics associated with the time of arrival and duration
by $\delta t_a$ and $\delta D$ respectively.
The metric $\delta t_a$ is simply the offset
\begin{eqnarray}
\delta t_a & = &  \widehat{t}_{a}-t_{a} \;,
\end{eqnarray}
where $\widehat{t}_{a}$ and $t_{a}$
are the times of arrival associated with 
the estimated and true amplitude envelopes
respectively.
Similarly, the metric $\delta D$ is  
\begin{eqnarray}
\delta D & = & \widehat{t}_{D} - t_{D}\;,
\end{eqnarray}
where $\widehat{t}_{D}$ and $t_{D}$ are the inter-quartile ranges associated with the 
estimated and true amplitude envelopes
respectively.

For $f(t)$, we adopt  the following  metric.
Let $\mathcal{F}$ be the set of time samples within the start and stop times of the true
signal. Let $\widehat{f}$ and $\overline{f}$ be the estimated and true instantaneous frequency sequences  respectively. Note that the set of time instants over which each is supported will not be identical in general. Define 
\begin{eqnarray}
\mathcal{G}(\epsilon_f) & = & \left\{ i \;|\; t_i\in \mathcal{F}, \left|[\overline{f}]_i - [\widehat{f}]_i\right|\leq \epsilon_f  \right\}\;.
\label{matchfreqset}
\end{eqnarray}
In words,
$\mathcal{G}(\epsilon_f)$ is that part of the true signal
where the estimated and 
true instantaneous frequencies differ by less than $\pm\epsilon_f$. 
The metric is then
defined as,
\begin{eqnarray}
\rho^2(\epsilon_f) & = & \frac{\sum_{i \in \mathcal{G}(\epsilon_f)} a_i^2}{\sum_{i \in \mathcal{F}} a_i^2}\;.
\label{eq:instf_metric}
\end{eqnarray}
The numerator is the squared norm of the true amplitude envelope, $\overline{a}$,  restricted to the samples in the set $\mathcal{G}(\epsilon_f)$. 
The denominator is the squared norm of the full $\overline{a}$. 

The metric $\rho(\epsilon_f)$ takes
account of the fact that the error
in frequency estimation can be
expected to be larger where the true signal amplitude is  weaker. Thus, we
must somehow weight the error by the instantaneous amplitude of the signal
before combining them. 
However, a straightforward 
weighted average of  $\widehat{f}-\overline{f}$, with the weight given by 
$p_a(t)$, is not found to perform well. This is because SEECR does not put any constraint on how 
fast $f(t)$ can vary and this allows the estimated frequency to change rapidly near the 
beginning and end of a signal where its true amplitude is small (or zero). (This effect is visible as a flaring of the estimates in Fig.~\ref{CC_instf_dist} 
around the end of the signal.) The resulting 
errors turn out to be too large to be compensated by the 
decaying amplitude envelope near these locations.  
By confining our attention 
to the interval $\mathcal{G}(\epsilon_f)$, where the estimated and true instantaneous 
frequencies agree well, and constructing
the metric out of the amplitude 
envelope, 
we cut out these spurious end effects and  fold in the required weighting at the same time.

\begin{figure}
    \includegraphics[scale=0.42]{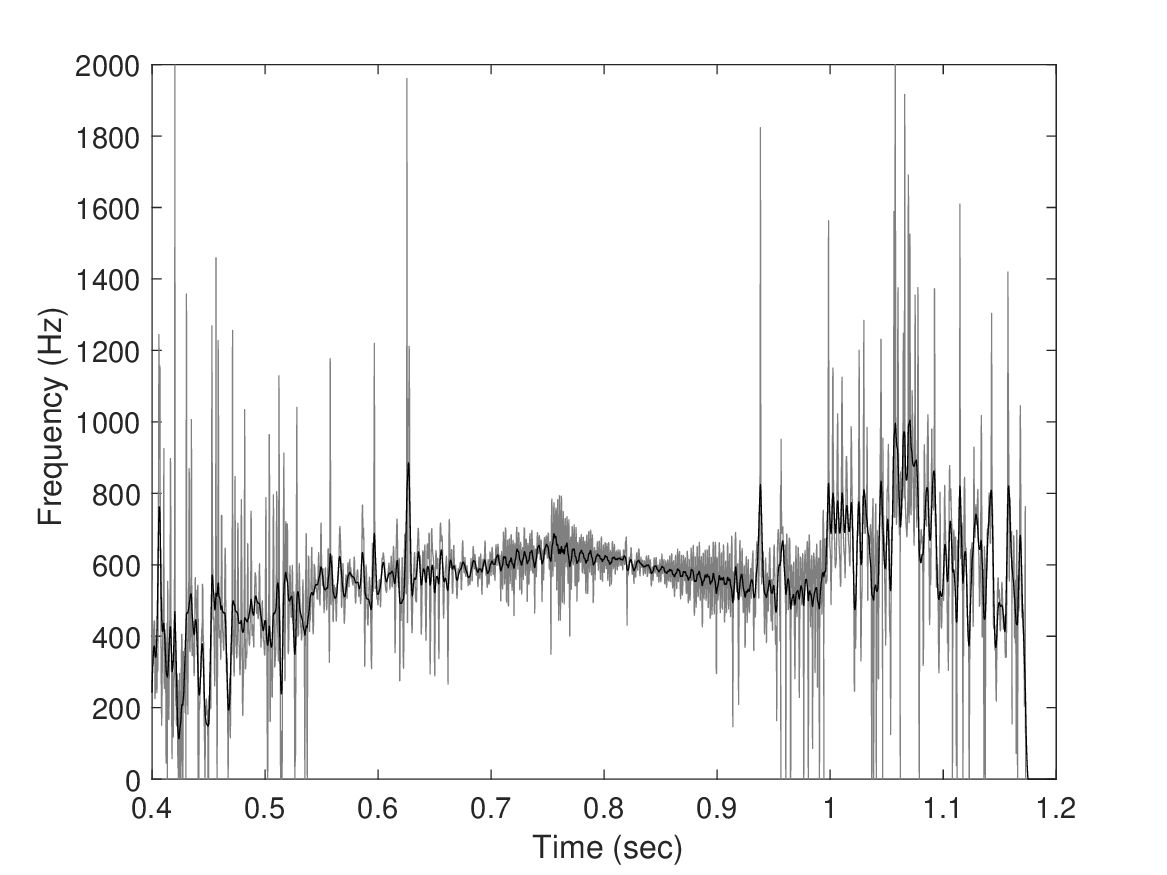}
    \caption{
    The instantaneous frequency, $\overline{f}$,
    of the s11WW signal (in gray) and its running average (in black). 
     $\overline{f}$ is obtained
     from the analytic form of the signal
     as described in 
     Sec.~\ref{sec:estimQuant}. The large spikes in $\overline{f}$, that cross a 
     band of $\approx \pm 300$~Hz around the running average, 
     are artifacts of this process and 
     should be ignored. The running average is computed 
     over a block of 10 samples. The slope of the trend changes from positive to negative
     somewhere in the interval $[0.75, 0.77]$~sec. Over this interval, the sample standard 
     deviation of $\overline{f}$ is $98.3$~Hz. 
    \label{s11WW_ft_ftrunningmean_fig}}
\end{figure}

While $\rho(\epsilon_f)$ as defined above 
is appropriate for a smoothly evolving 
instantaneous frequency, 
it needs to be modified for signals where this 
is not true.
As can be seen from 
Fig.~\ref{s11WW_ft_ftrunningmean_fig}, the s11WW signal presents such a situation, where,
In addition to an underlying trend,
a fair amount of 
scatter (excluding the spurious 
spikes) is
evident in the true instantaneous frequency. 
The trend can be elucidated by taking a
running average, which is also shown in the figure. 
The scatter must be accounted for 
when comparing estimated and true instantaneous frequencies 
because no semi-parametric method, such as 
SEECR, can hope to match the scatter in detail without having a degree of freedom
that is so large as to make it practically
useless.

For the s11WW signal, therefore, $\rho(\epsilon_f)$ is calculated with $\overline{f}$
replaced in Eq.~(\ref{matchfreqset}) by its running average. It should be noted that 
setting $\epsilon_f$ to be less than the standard deviation of
the running average itself will again show up  
as an apparent loss in performance. The running average used here is computed over 
a block of 10 samples, and given that the standard deviation of $f_i$
around the running average is $\approx 100$~Hz,  the standard deviation of the running average
itself is $\approx 30$~Hz.

\subsubsection{Metric distributions}
\label{metric_distributions}

\begin{figure}
    \includegraphics[scale=0.42]{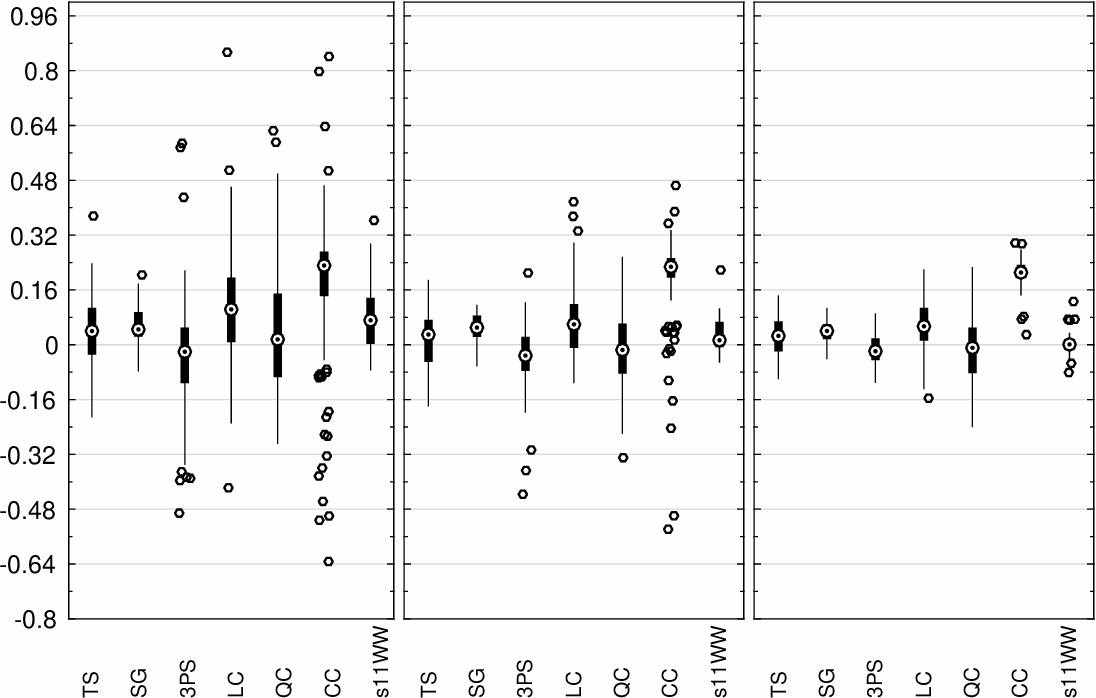}
    \caption{
        Box-and-whisker plots of the metric $\delta t_a$. 
    Each box-and whisker summarizes the sampling distribution, as described 
    in Sec.~\ref{metric_distributions}, of $\delta t_a$ for one signal 
    and one SNR. The name of the signal is shown on the X-axis.
    The box-and-whisker plots corresponding to the same SNR are grouped in
    one planel. From left to right, the panels correspond to SNR values of $[10, 12, 15]$
    respectively.
    \label{delta_ta_medianbased_fig}}
\end{figure}

\begin{figure}
    \includegraphics[scale=0.42]{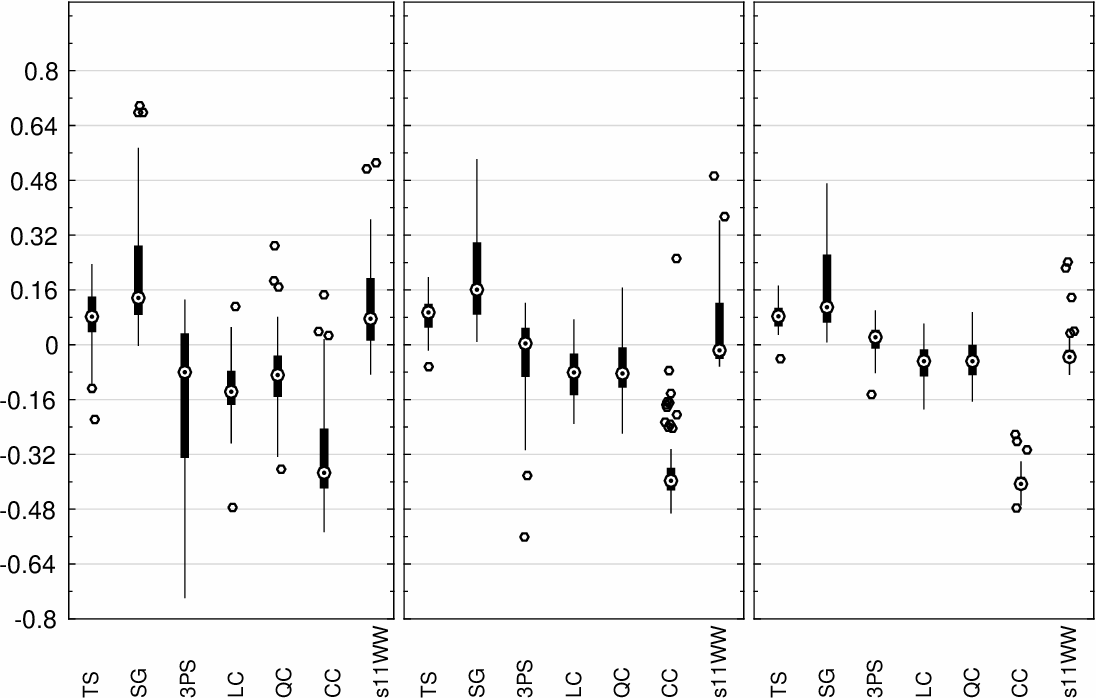}
    \caption{
        Box-and-whisker plots of the metric $\delta D$. 
    Each box-and whisker summarizes the sampling distribution, as described 
    in Sec.~\ref{metric_distributions}, of $\delta D$ for one signal 
    and one SNR. The name of the signal is shown on the X-axis.
    The box-and-whisker plots corresponding to the same SNR are grouped in
    one panel. From left to right, the panels correspond to SNR values of $[10, 12, 15]$
    respectively.
    \label{fig:delta_D}}
\end{figure}

\begin{figure}
    \includegraphics[scale=0.42]{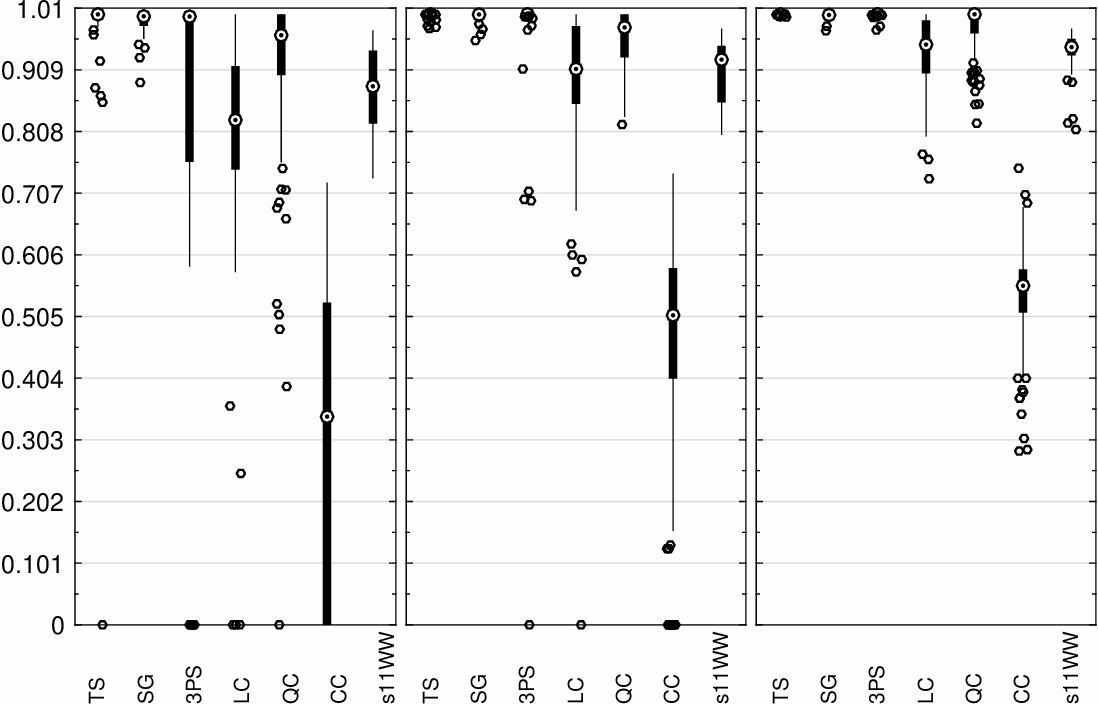}
    \caption{
        Box-and-whisker plots of the  metric $\rho(\epsilon_f)$ for $\epsilon_f = 64$~Hz. 
    Each box-and whisker summarizes the sampling distribution, as described 
    in Sec.~\ref{metric_distributions}, of $\rho(\epsilon_f)$ for one signal 
    and one SNR. The name of the signal is shown on the X-axis.
    The box-and-whisker plots corresponding to the same SNR are grouped in
    one panel. From left to right, the panels correspond to SNR values of $[10, 12, 15]$
    respectively.
    \label{instFreq_rhometric_ampnormratio_maxFreqTol_fig}}
\end{figure}

Fig.~\ref{delta_ta_medianbased_fig},
Fig.~\ref{fig:delta_D}, and
Fig.~\ref{instFreq_rhometric_ampnormratio_maxFreqTol_fig} 
summarize the
sampling distributions of $\delta t_a$, $\delta D$, and $\rho(\epsilon_f)$ respectively in the form of 
box-and-whisker plots. For each box, 
the `$\odot$' mark indicates the median
of the distribution, while the 
 bottom and top edges correspond to its $25^{\rm th}$ and $75^{\rm th}$ percentiles respectively. Thus, the length of a box 
corresponds to the IQR and contains $50\%$ of the probability. The whiskers (thin lines) 
extend to the extreme data points that are not outliers. 
A sample value is deemed to be an 
outlier if it is separated from the median by more than twice the IQR. (Outliers are shown as open circles
that are dithered horizontally by small amounts to aid visual clarity.)

For reference, the length of each whisker is $2.68\sigma$ for a Normal distribution having a standard deviation $\sigma$. 
As such,  
more than $99\%$ of the probability under a Normal distribution 
is contained between the ends of the two whiskers.
While this is also true for the observed distributions in general, there are some 
exceptions. The correct probability coverage in such cases can be obtained 
by simply counting the number of outliers in the plot and subtracting it
from the number of trials (see
Table~\ref{detprobtable} for the exact number of trials). 
 
The distributions of $\delta t_a$ and $\delta D$ show that reducing the errors 
in time of arrival and duration, if they are obtained from the estimated 
amplitude envelope, to levels where they are significantly smaller than the duration of 
the signals requires ${\rm SNR} \gtrapprox 15$.
The lowest error at this ${\rm SNR}$ is in the range $\pm 0.08$~sec, with a probability of $\gtrapprox 0.99$,
for the
 s11WW signal ($1$~sec duration). With the same probability, QC shows the broadest range for the error 
 at about $\pm 0.24$~sec.
 
While the time of arrival is generally estimated with negligible bias, it is significant 
for the CC signal due to its partial
reconstruction (see Fig.~\ref{CC_instf_dist}). However,
the bias is fairly independent of ${\rm SNR}$ and, hence, will not affect the 
offsets between the estimated times of arrivals for CC signals in a network of detectors.
Excluding CC, the largest range ($\approx 0.99$ probability) in time of arrival error at ${\rm SNR}=10$, which occurs
for the QC signal, is $\pm 0.5$~sec.

The bias in duration estimation, on the other hand, is non-negligible for several
signals even at ${\rm SNR}=15$. The anomaly in the duration estimation is the SG signal, for which
the error has a distinctly asymmetrical distribution around the median. This is because the estimated amplitude envelope for 
this signal has a peak that is well localized around that of the true signal, as evident from its 
$\delta t_a$ distribution, but it is biased away from having a 
symmetrical shape around the peak. This illustrates the problem, mentioned earlier, with using the start and 
stop times of the estimated amplitude envelope directly for deriving  duration and time of arrival.

From Fig.~\ref{instFreq_rhometric_ampnormratio_maxFreqTol_fig}, we see that
with a tolerance of $\epsilon_f= 64$~Hz in frequency estimation error,
SEECR is able to 
recover $\gtrapprox 60\%$ of the frequency evolution at the lowest ${\rm SNR}$ with  
$\gtrapprox 0.99$ probability.
The only exception is the CC signal
 and the reason is again its partial reconstruction. Excluding this signal, the whiskers
 of all the distributions  lie above $\approx 70\%$ at ${\rm SNR}=12$ and $\approx 80\%$ at
 ${\rm SNR}=15$. 
 
\begin{figure}
    \includegraphics[scale=0.42]{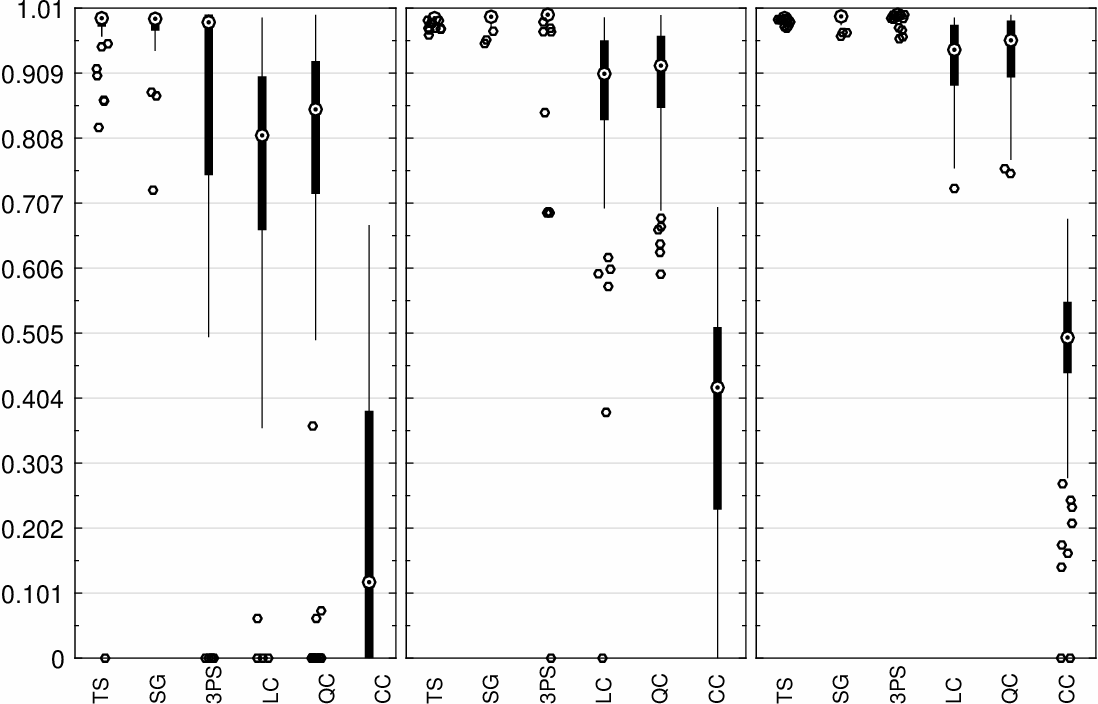}
    \caption{
        Box-and-whisker plots of the  metric $\rho(\epsilon_f)$ for $\epsilon_f = 4$~Hz. 
    Each box-and whisker summarizes the sampling distribution, as described 
    in Sec.~\ref{metric_distributions}, of $\rho(\epsilon_f)$ for one signal 
    and one SNR. The name of the signal is shown on the X-axis.
    The box-and-whisker plots corresponding to the same SNR are grouped in
    one panel. From left to right, the panels correspond to SNR values of $[10, 12, 15]$
    respectively.
    \label{instFreq_rhometric_ampnormratio_maxFreq4_fig}}
\end{figure}

Fig.~\ref{instFreq_rhometric_ampnormratio_maxFreq4_fig} shows the distribution
of $\rho(\epsilon_f)$ for a much tighter tolerance of
$\epsilon_f = 4$~Hz. 
The distribution for the s11WW signal is not included in this figure because 
$\epsilon_f$ is  smaller than the standard deviation of the running average of its 
instantaneous frequency (see Sec.~\ref{sec:estimation_metrics}). 
 The changes in the distributions of $\rho(\epsilon_f)$ are relatively small
  for all the other signals and $\gtrapprox 75\%$ of frequency evolution
  is still recovered at ${\rm SNR}= 15$ with  $\gtrapprox 0.99$ probability.


\subsection{GW150914 analysis}
\label{seecr_detestperf_150914}
\begin{figure}
    \includegraphics[scale=0.38]{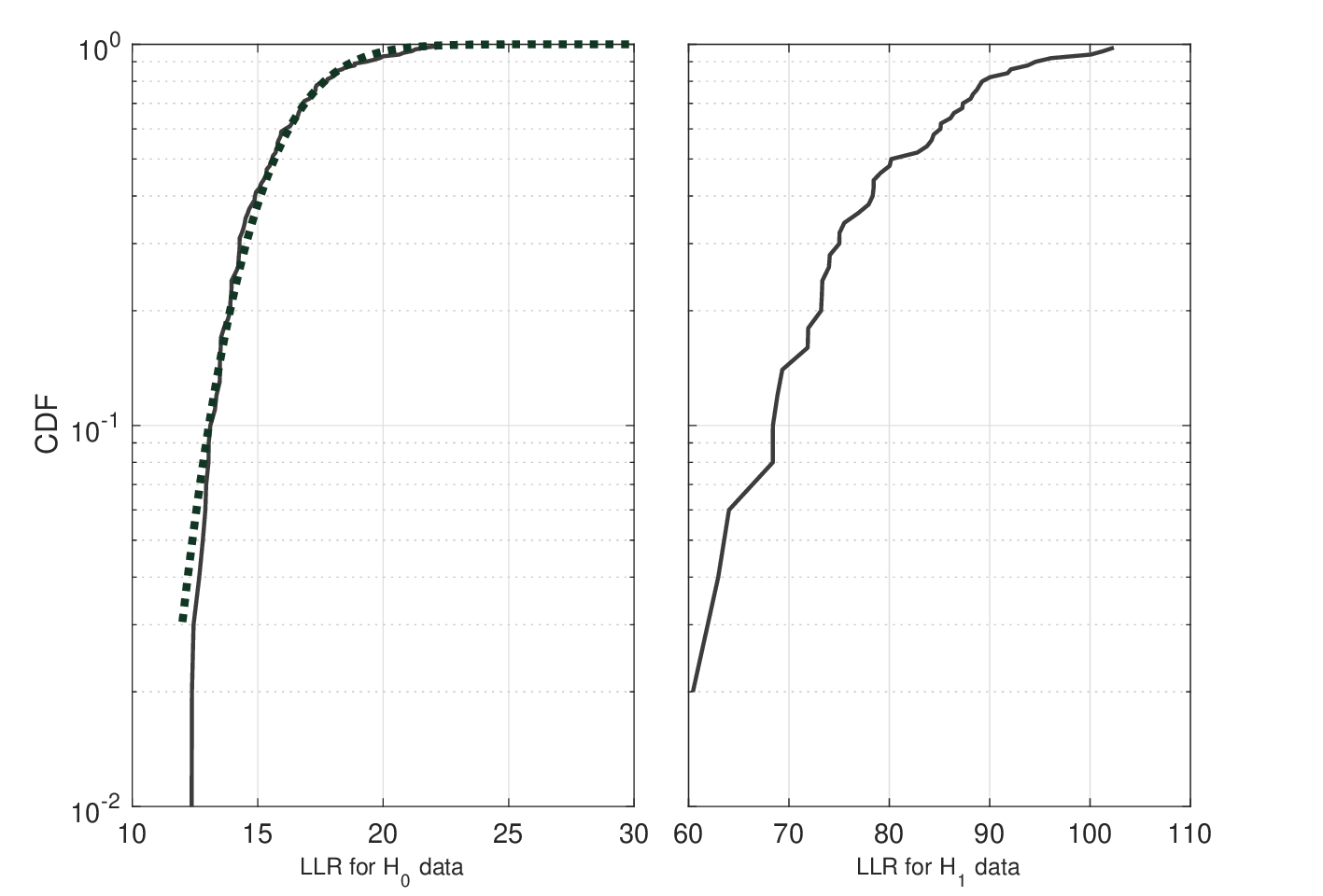}
    \caption{
        Estimated cumulative distribution functions (CDFs) of the LLR
        obtained from the 
        SEECR analysis of GW150914 data. The CDF on 
        the left (solid curve) is estimated from noise-only data (generated 
        independently of GW150914 data) while the one on the right corresponds
        to the real GW150914 data with added pseudo-random noise.  Also shown on the left (dotted curve) is the CDF of the best lognormal fit. 
    \label{fig:GW150914_cdf_H0_H1}}
\end{figure}

As 
described in Sec.~\ref{realsignals}, pseudo-random noise was added 
to GW150914 data to reduce the observed ${\rm SNR}$ of the signal by a factor of 2.
Fig.~\ref{fig:GW150914_cdf_H0_H1} 
shows the cumulative distribution function of the LLR [Eq.~(\ref{eq:detstatistics})] under $H_0$, along with a lognormal fit, and $H_1$. The two distributions do not overlap. Based on the lognormal fit, SEECR
can detect a signal like
GW150914 at an ${\rm SNR}=10$ with a probability of unity even 
at a false alarm probability of $2\times 10^{-16}$ (corresponding to a threshold of ${\rm LLR}=50$).  

\begin{figure}
    \includegraphics[scale=0.3]{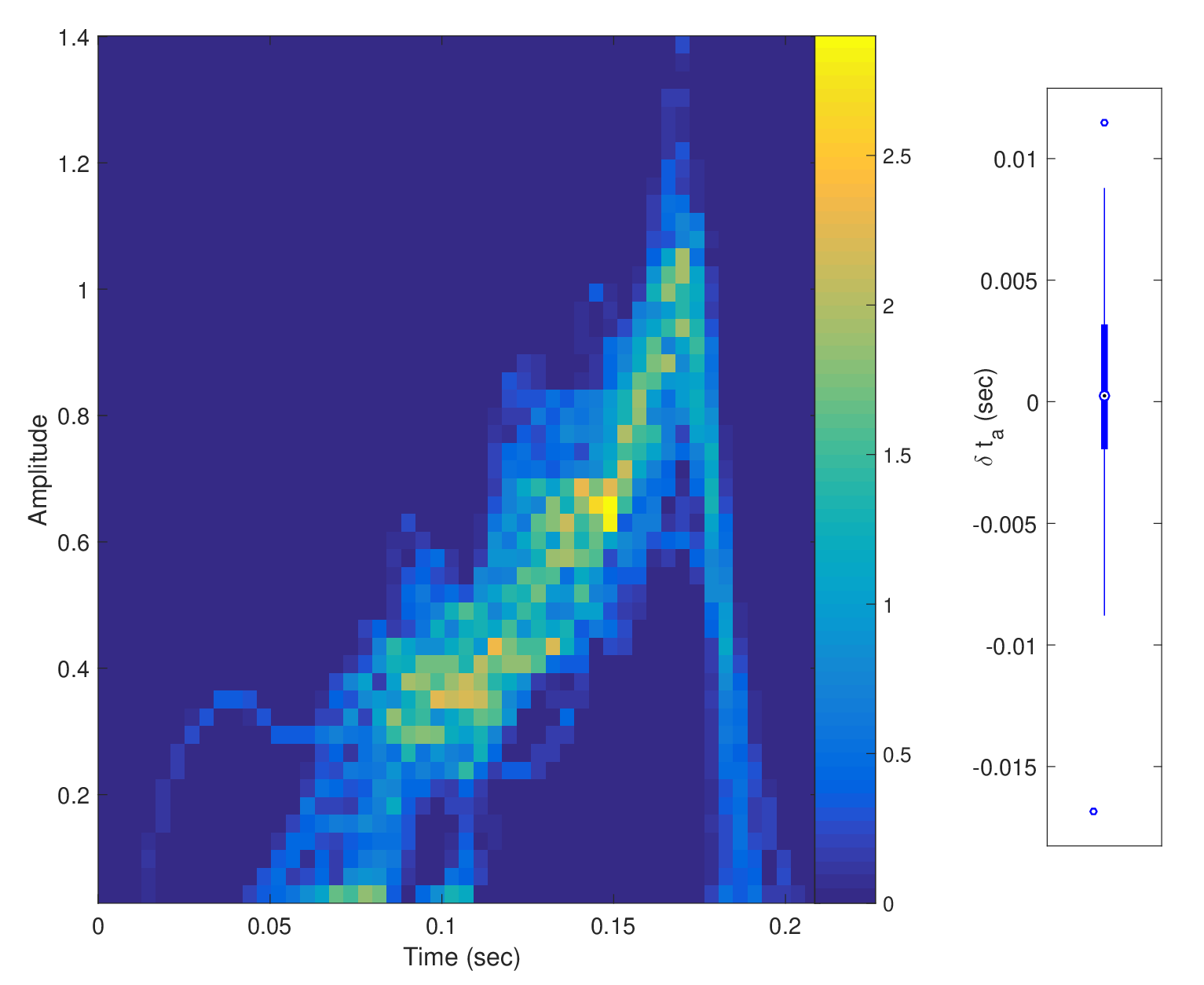}
    \caption{ The 2D histogram of estimated amplitude envelopes (left panel) using 50 data realizations of GW150914 data with added pseudo-random noise. The histogram 
     is constructed by plotting all the estimated amplitude envelopes and counting the number of plotted
     points in a regular grid of 2D bins. There are 50 bins along each dimension. The counts have been normalized by the number of realizations used. The distribution of the metric, $\delta t_a$, is shown as a box-and-whisker plot (right panel). The true time of arrival was taken to be the one associated with the signal 
     estimated by SEECR from the original GW150914 data. 
    \label{fig:gw150914_2dhist_ampenvlp}}
\end{figure}

\begin{figure}
    \includegraphics[scale=0.3]{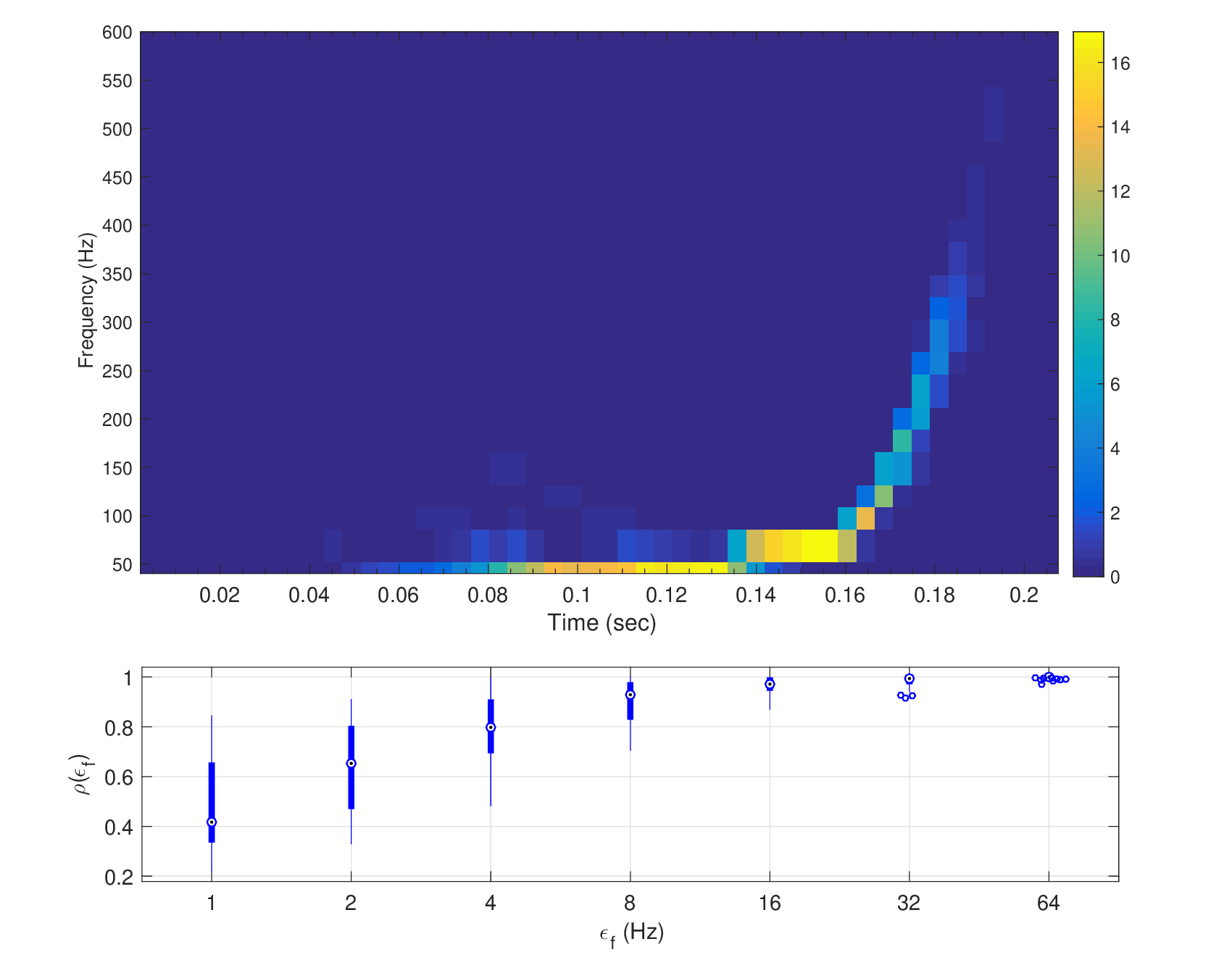}
    \caption{ The 2D histogram of estimated instantaneous frequencies (top panel) using 50 data realizations of GW150914 data with added pseudo-random noise. The histogram 
     is constructed by plotting all the estimated instantaneous frequencies and counting the number of plotted
     points in a regular grid of 2D bins. There are 50 bins along each dimension. The counts have been normalized by the number of realizations used. The distribution of the metric, $\rho(\epsilon_f)$, is shown as a box-and-whisker plots (bottom panel) for $\epsilon_f \in \{1,2,4,8,16,32,64\}$~Hz. For the calculation of 
     $\rho(\epsilon_f)$, the true amplitude envelope 
     was taken to be that of the signal estimated by SEECR from the original GW150914 data. 
    \label{fig:gw150914_2dhist_instFreq}}
\end{figure}

Fig.~\ref{fig:gw150914_2dhist_ampenvlp} and Fig.~\ref{fig:gw150914_2dhist_instFreq} show the 2D histograms, following the construction
described in Fig.~\ref{CC_instf_dist}, of all the estimated 
amplitude envelopes and instantaneous frequencies
respectively, along with box-and-whisker plots of the metrics $\delta t_a$ and $\rho(\epsilon_f)$. 
Comparison of the 2D histograms clearly illustrates the discussion in
Sec.~\ref{outermin}
that the 
estimation error for the amplitude envelope of a chirp is significantly higher than 
that for its instantaneous frequency.

From the $\delta t_a$ distribution, we conclude
that,  with a probability of about $\approx 0.5$ and $\approx 0.99$ respectively,
SEECR was able to pin down the time of arrival of the signal to within
about $\pm 2.5$~msec and $\pm 10$~msec. This is a vast improvement
over the situation seen in Sec.~\ref{metric_distributions}
for the case of long duration signals. 
The $\rho(\epsilon_f)$ distribution, on 
the other hand, shows a worsening 
relative to the longer duration signals. For example,
compared to the SG signal at ${\rm SNR=10}$, which 
shows $\gtrapprox 90\%$ recovery with $0.99$ probability
for $\epsilon_f = 4$~Hz, the same
performance for GW150914 requires an error tolerance of 
$\epsilon_f \gtrapprox  16$~Hz .


\section{Comparison with time-frequency clustering}
\label{seecr_tf_detestperf}

All of the  principal search algorithms used in LIGO for GW burst search~\cite{PhysRevD.93.042004,cornish2015bayeswave,2017PhRvD..95j4046L} use
some form of time-frequency (or time-scale) clustering. 
It is assumed that the presence
of a signal in noisy data produces areas of locally high power,
or {\em clusters},
in the time-frequency plane.  Depending on the properties used for
distinguishing between signal and noise induced clusters, there is 
 a wide variation in how clustering is implemented, ranging from
a nearest neighbor based approach~\cite{sylvestre2002time} to a proximity prior~\cite{cornish2015bayeswave}.

For a given ${\rm SNR}$, the
sensitivity of any clustering based method is naturally lower for signals that 
do not produce strong clusters. This is a particularly relevant issue
for chirps
since they spread their total energy over an extended track.  
Therefore, it is interesting to 
compare the performance of  SEECR with time-frequency clustering.

Since 
a full-fledged comparison with the search methods used in LIGO is outside
the scope of this paper,
 we construct an ad hoc clustering based search method that is simpler but, at the same time, captures the principal features of
 clustering used in the more sophisticated methods. 
 We refer the reader to Appendix~\ref{app:clusteringsearch} for a description of  
the clustering based search method. Here, we focus entirely on the results 
obtained with this method and its comparison with SEECR.

To quantify the performance of the clustering based search method, we 
generate data realizations  in exactly the same way as described in Sec.~\ref{sims}.
However, due to the use of multi-resolution analysis (see Appendix~\ref{app:clusteringsearch}),  the overall FAR is split across the different resolution levels and, consequently, a much larger number of data realizations is required to reduce sampling errors.
Consequently, we generate $10^4$ and $10^3$  
$H_0$ and $H_1$ data realizations respectively. For the same reason, we 
only compare the clustering based method and SEECR at the larger 
FAR of $10^{-3}$~events/sec.

\begin{figure}
    \includegraphics[scale=0.47]{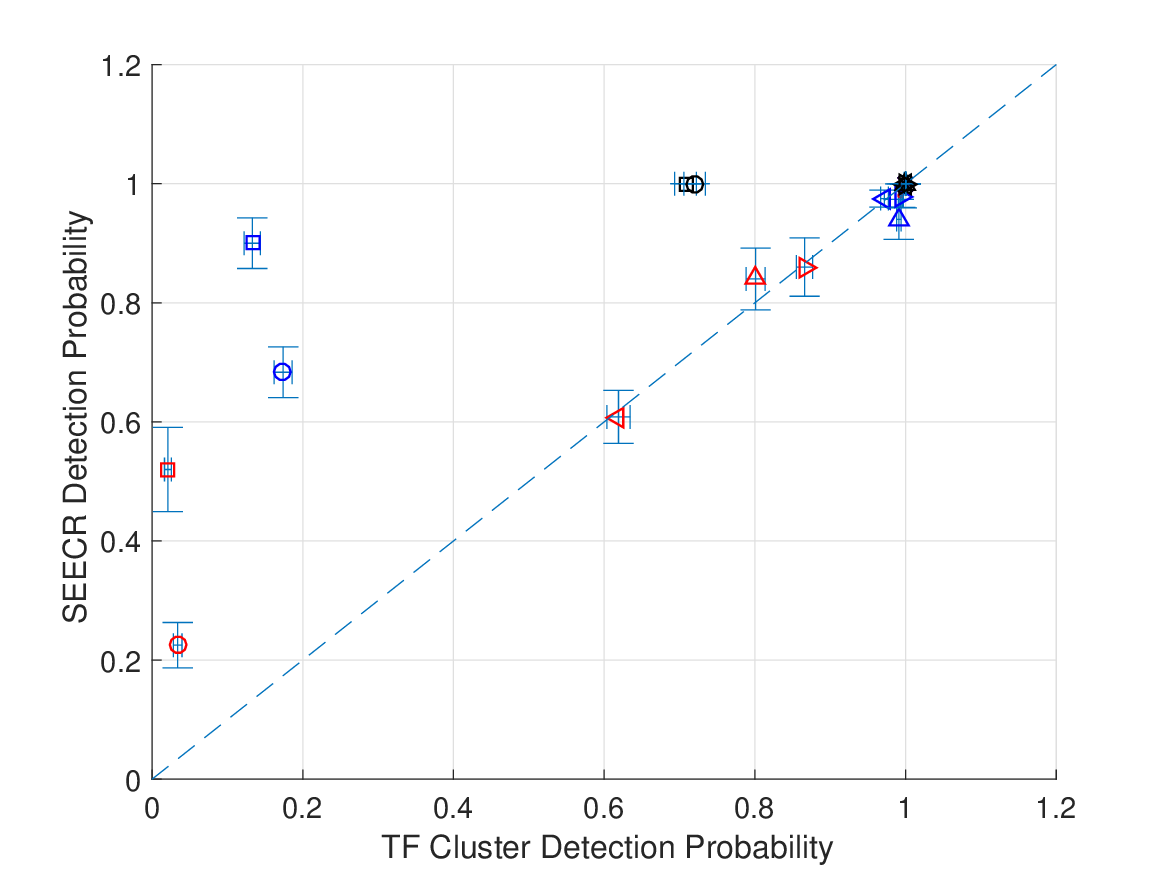}
    \caption{ Scatterplot of detection probability attained by Time-frequency 
    clustering (x-axis) and SEECR (y-axis) for the simulated signals (see Sec.~\ref{sims}) and ${\rm SNR} = 10, 12, 15$. (The axes have been extended beyond a probability of unity for clarity.) The marker shapes correspond to the
    signal waveforms as follows. TS ($\bullet$), SG ($\ast$), 3PS ($\triangle$), LC ($\Box$), QC ($\triangleleft$), CC ($\circ$), s11WW ($\triangleright$). The color of a marker indicates the ${\rm SNR}$ with the correspondence: $10$ (red), $12$ (blue), and $15$ (black). The error bars in each direction correspond to the respective $1\sigma$ intervals [c.f., Eq.~(\ref{eq:errorbar_prob})].
    \label{fig:Seecr_TFClstrs_detprob}}
\end{figure}

Fig.~\ref{fig:Seecr_TFClstrs_detprob} shows a scatterplot of the detection probability attained by SEECR (from Table~\ref{detprobtable}) and
the clustering based method across all signals and ${\rm SNR}$ values.
We see from the points that are far away from the line of equal detection probabilities that the performance of SEECR is significantly better than clustering for the CC and LC signals. For the remaining signals, the two have essentally
the same performance. 

At ${\rm SNR}=15$, the detection probabilities
attained by clustering for the LC and CC signals  are
$0.71\pm 0.014$ and $0.72\pm 0.014$ respectively while they are unity for 
SEECR in both cases. The performance of clustering worsens rapidly for these 
signals as ${\rm SNR}$ is reduced, with the detection probabilities at ${\rm SNR} = 12$ being $0.133\pm 0.011$ and $0.174\pm 0.011$ for LC and CC respectively. (The corresponding probabilities are $0.90\pm0.04$ and $0.68\pm0.04$ for SEECR.)
While a reduction in performance of clustering is expected, due to
the spreading of signal power across a track, 
the extent to which it degrades for a simple signal such as LC is
quite surprising.

\begin{figure}
    \includegraphics[scale=0.29]{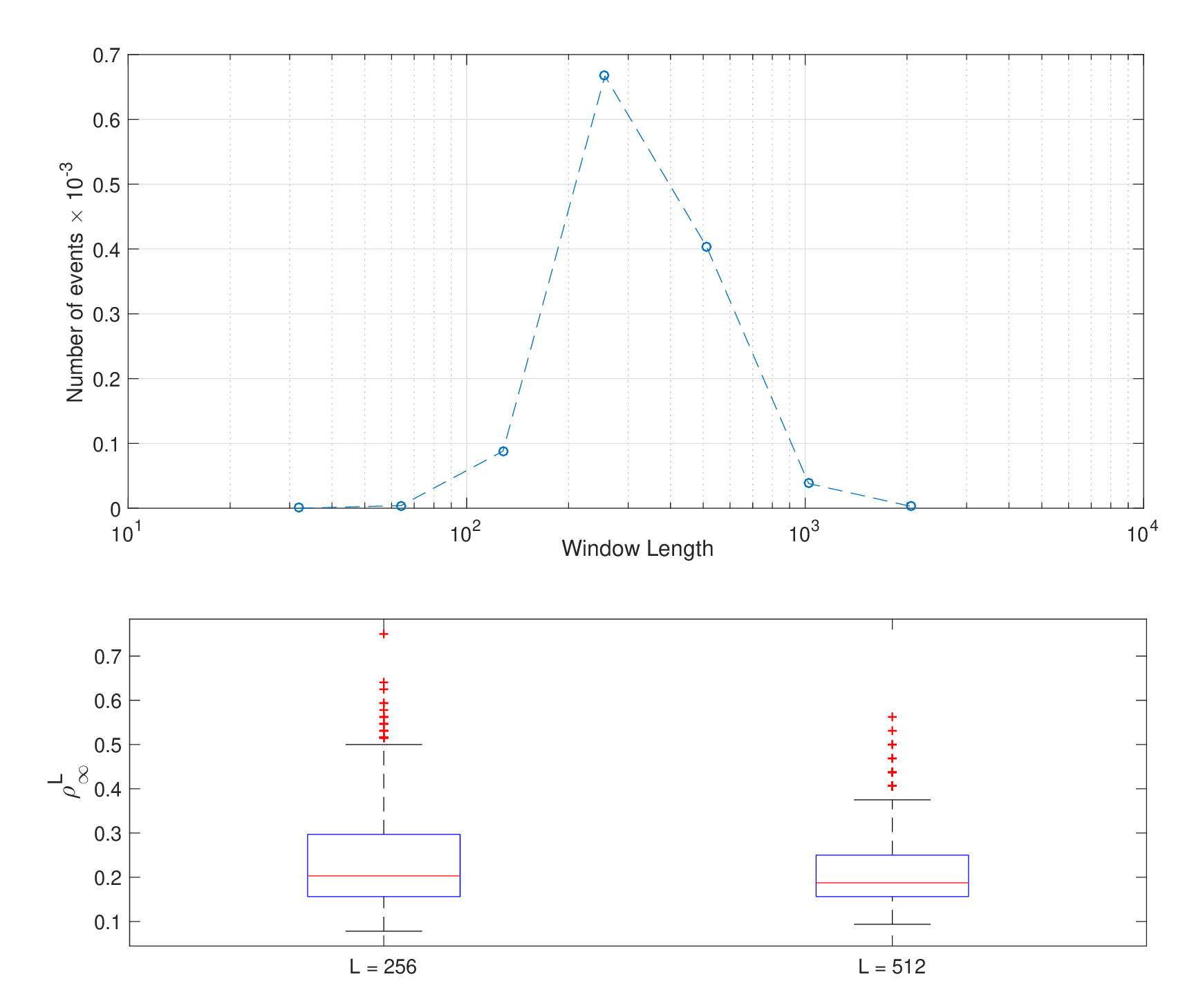}
    \caption{The top panel shows the number of time-frequency events found, as a fraction of the number of data realizations ($=1000$), for each of the window lengths, $L=2^n$, $n = 5, 6, \ldots, 11$, used in the clustering based search method. Each data realization contains the LC signal with an ${\rm SNR}=15$.
    (The dashed line is included as a visual aid only, with the actual data points shown as open circles.)
    The bottom panel shows 
     the distribution of $\rho_\infty^L$ for $L= 256$ and $L=512$, the window lengths that produce the bulk of detected events, 
     in the form of box-and-whisker plots. In these
     plots, the median is indicated by the red line in each box. The edges of 
     each box and the size of the whisker carry the same meaning as described in Sec.~\ref{metric_distributions}. Outliers are denoted by the `$+$' marker.
    \label{fig:TFclstrs_linChirp_eventStats}}
\end{figure}

While clustering is a detection, not an estimation,
method,   estimation is possible as a follow up step to clustering based 
detections. However, if the estimation algorithm
focuses on only the time-frequency regions 
identified as significant by the clustering step,
the errors in the estimation can become quite large. This is evident from Fig.~\ref{fig:TFclstrs_linChirp_eventStats} 
where we have taken the case of 
data realizations containing the LC signal at ${\rm SNR}=15$ and analyzed the 
associated time-frequency events as described below. (See Appendix~\ref{app:clustering} for
the definition of a time-frequency event.)

Let $C_0^L$ be the set of spectrogram columns constituting the support of 
the true signal for window length $L$, and let $C^L$ be the set of 
columns constituting an event. The ratio $n(C^L\cap C_0^L)/n(C_0^L)$, where $n(A)$
is the cardinality of a set $A$, 
is a simple measure of how well clustering can indicate the time-frequency region for
follow up analysis by estimation algorithms. For a signal such as LC that has 
a constant amplitude envelope over its entire duration, this ratio is equivalent
to the metric $\rho(\epsilon_f)$ defined in Eq.~(\ref{eq:instf_metric}) but with
$\epsilon_f$ set to be the entire frequency range of the spectrogram. To indicate this 
connection, we denote
the ratio above as $\rho_\infty^L$.

As can be seen from the box-and-whisker plots in Fig.~\ref{fig:TFclstrs_linChirp_eventStats}, for the window lengths $L=256$ and $L=512$ that 
produce the bulk of the detected events, 
 $\rho_\infty^L$ is $\lessapprox 0.3$ with a probability of $0.75$ and, consequently, clustering 
 flags $\lessapprox 30\%$ of the region of the time-frequency plane containing the true 
 signal. (This fraction would be reduced further if the error in frequency estimation is also taken into account.) In contrast, we see from Fig.~\ref{instFreq_rhometric_ampnormratio_maxFreq4_fig} that
 SEECR recovers $\approx 90\%$
of the LC signal at the same ${\rm SNR}$ and probability with a frequency estimation 
error of $\pm 4$~Hz.

\section{Comparison with Track-Search and CPP}
\label{seecr_cpp_trksrch_detestperf}

The GW150914 system, with
each of its components having a measured mass  of $\approx 30$~$M_\odot$, falls within  the range of simulated BBH signals used in~\cite{1999PhRvD..60j2001A}
for analyzing the performance of Track-Search. 
Hence, the results in Sec.~\ref{seecr_detestperf_150914} pertaining to the 
analysis of GW150914 data at an observed ${\rm SNR}=10$
 can be used to
compare the performance of SEECR with that of Track-Search. 

The analysis in~\cite{1999PhRvD..60j2001A} uses a false alarm probability of $3.4\times 10^{-5}$ for segments that are $0.415$~sec long, sampled at a frequency of
$9868.42$~Hz.  This corresponds to a FAR of $8.2\times 10^{-5}$~events/sec.
Assuming that the FAR scales linearly with the frequency search range, 
and that the range used in Track-Search 
extended to the Nyquist frequency of $4934.21$~Hz, 
the equivalent FAR for SEECR is $(450/4934.21)\times
8.2\times 10^{-5} = 7.5\times 10^{-6}$~events/sec. Here, we have used the actual 
bandwidth of $[0,450]$~Hz of the GW150914 data even though SEECR was run with a 
frequency search range that extends to $2000$~Hz. Finally, translating
this FAR back to false alarm probability for the GW150914 data segment
length of $0.21$~sec, we get $1.6\times 10^{-6}$. 

Based on the log-normal fit in Fig.~\ref{fig:GW150914_cdf_H0_H1}, 
the false alarm probability
derived above
corresponds to a threshold of $30.4$ on the LLR statistic. At this
threshold, the detection probability attained by SEECR is unity. 
The detection probability for Track-Search can be read off from Fig.~5 of~\cite{1999PhRvD..60j2001A} to be $\approx 0.8$ for a system with a total mass of 
$60$~$M_\odot$ at ${\rm SNR}=10$. 
With the caveat that a proper comparison requires
analysis of the  
same data realizations with compatible search parameter settings and a wide
range of waveform morphologies, 
we find that SEECR has a performance that is comparable
to or better than that of Track-Search.

For comparing SEECR with CPP, we use a different
simulation setup than the one in Sec.~\ref{simwaveforms}. This 
was necessitated by the high  Random Access Memory (RAM) 
requirement of the public domain CPP code (ChirpLab), 
which results in very large
execution times when applied to the data realizations in Sec.~\ref{sims}. 
Each data realization is now shorter, with a duration of $0.5$~sec at a sampling frequency of $2048$~Hz, leading to $1024$
samples per realization. 

We generate $500$ realizations of
$H_0$ data following the same noise model as in Sec.~\ref{sims}. For $H_1$ data, we use a Newtonian inspiral signal that  starts at $0.1$~sec and lasts $0.24$~sec. This signal corresponds to an 
equal mass binary with a total mass of $45.0$~$M_\odot$ and a lower frequency cutoff of $40$~Hz. To stay within the range 
of signal strengths used in~\cite{candes2008gravitational}, 
the signal is normalized to have a value of $0.25$ for the SNR as
defined in~\cite{candes2008gravitational}, which
corresponds to  ${\rm SNR} = 5.5$ as per the definition in this paper.
(The latter is a factor of $\sqrt{N}$ higher than the former,
where $N=492$ is the number of samples in the signal
waveform.)

We use the Best Path statistic~\cite{candes2008gravitational}
for path lengths
$[1, 2, 4, 8, 16]$ and the routines  provided in ChirpLab for estimating 
detection probability.  
At
a false alarm probability of $0.05$, which is the fiducial
value used in~\cite{candes2008gravitational},
CPP is found to attain a detection probability of 
$0.242$. Within sampling error, and for the same false alarm probability,
SEECR gives a nearly identical detection probability
of $0.25$. Thus, CPP and SEECR 
seem to be comparable in performance.

The CPP algorithm  assumes that the signal waveform occupies the whole 
of the data segment being analyzed. Although this condition is violated by the $H_1$ data
 described above, it provides a more realistic test since
the true duration of an unmodeled signal is  unknown by definition. 
That said, 
a version of CPP that is not limited by the above assumption should be used in future studies
for a fair comparison.

 
 \section{Conclusions}
 \label{conclusions}
We have presented a novel algorithm, called SEECR, for the 
 detection and estimation of unmodeled transient chirp signals.
 The algorithm makes no assumptions about the time evolution 
 of the amplitude envelope, $a(t)$, or the instantaneous frequency, $f(t)$, of a chirp
 signal except that they are smooth. The smoothness requirement
 is incorporated by modeling $a(t)$ and $f(t)$ with mutually independent splines.
An important feature of SEECR is its small number of free parameters.
Given enough computing power, these can be reduced to  
just two, namely, the maximum number of breakpoints to use for the two splines.

  SEECR was tested on 7 chirp signals spanning a wide
  range of amplitude and frequency evolution morphology, 
  and found to achieve a 
 detection probability $\gtrapprox 0.5$ in the low ${\rm SNR}$ range ($10\leq {\rm SNR}\leq 12$)
 at a FAR between $10^{-3}$~events/sec and $10^{-4}$~events/sec. 
 Hence, it is capable of achieving good sensitivity at astrophysically realistic
 signal strengths. 
 
 In terms of estimation, the instantaneous frequency of a signal
 is estimated much better, as expected, than the amplitude envelope. For example, 
 excluding the CC signal due to its partial reconstruction, more than $60\%$ of the 
 instantaneous frequency evolution of a signal can
 be recovered with an error of
 $\pm 64$~Hz at ${\rm SNR}=10$.  
 
 Errors in the time of arrival were found to 
 depend strongly on the true signal duration. For
 example, the smallest error range found is about $\pm 80$~msec for the
 s11WW signal ($1$~sec duration)
 at ${\rm SNR}=15$ but it reaches the $\pm 10$~msec
 level, comparable to the maximum light travel time between the two LIGO detectors,
 for GW150914 ($0.2$~sec duration) at a lower ${\rm SNR}$ of $10$.
 
 If SEECR is used in a temporal   
 coincidence scheme
 across two GW detectors with an acceptance window of $\pm 0.5$~sec,
 which is the largest error range at the lowest SNR,
 the coincidence FAR becomes $(10^{-4})^2$~events/sec, or $1$~event in $3.17$~years, for a single detector FAR of $10^{-4}$~events/sec. At ${\rm SNR}=12$, the lowest
 corresponding two-detector coincidence detection probability, excluding the CC signal, 
 is $0.84^2 \approx 0.7$  for the LC signal.
 
 The coincidence FAR
 can be reduced substantially if instead of a temporal
  scheme, coincidence is 
 imposed on the estimated frequency evolution. 
 However, we did not explore coincidence schemes further 
 in this paper because it 
 is not the optimal way to utilize
 multiple GW detectors. 
 The  proper generalization of SEECR, which is a major future direction for its evolution, is coherent network analysis where each of the two GW polarizations is an independent instance of the single-detector signal model used in this paper. An early step in this direction is reported in~\cite{calvin-siuro} for the much simplified case where each GW polarization waveform itself is assumed to be a spline.

 Based on an ad hoc time-frequency clustering method, 
 we found that SEECR  
 significantly outperforms a clustering based search 
 for some of the signal waveforms (CC and LC).
 At a FAR of $10^{-3}$~events/sec and ${\rm SNR}=12$, 
 the clustering based method
 could only achieve detection probabilities in the 
  $[0.133\pm 0.011, 0.174\pm 0.011]$ range while SEECR 
  achieved $[0.90\pm 0.04,0.68\pm 0.04]$.
Our clustering method fully incorporates multi-resolution 
analysis, which is the main driver of performance for such 
methods. Hence, we do not expect a significantly different 
outcome for more sophisticated approaches to the production of
time-frequency clusters. 

 Since clustering is a key component of the burst search methods used by LIGO, 
 SEECR can complement current searches by
 extending their coverage of 
 GW waveform morphologies.
 We also compared SEECR to Track-Search and CPP and found
 that it is comparable in performance to these methods. 
 
 The metrics proposed here to quantify the estimation performance of SEECR can prove useful for a comparative study of algorithms that target long duration ($\gtrapprox 1$~sec) chirp signals. 
 Similarly, the set of waveforms used here
 can be serve as a  
 benchmarking testbed. 
 
 \begin{acknowledgements}
 We thank Prof.~E.~Cand\`es for providing the ChirpLab code for CPP, and Prof.~I.~Pinto and
 Prof.~L.~Troiano for helpful discussions. 
 This work was supported by National Science Foundation Grant No. PHY-1505861.
 We acknowledge the Texas Advanced Computing Center (TACC) at The University of Texas at Austin for providing HPC resources that have contributed to the research results reported within this paper. URL: http://www.tacc.utexas.edu
 This research has made use of data obtained from the LIGO Open Science Center (https://losc.ligo.org), a service of LIGO Laboratory and the LIGO Scientific Collaboration. LIGO is funded by the U.S. National Science Foundation.
 \end{acknowledgements}
 
 \appendix
 \section{B-spline functions}
 \label{app_bsplines}
 A spline is a piecewise polynomial function defined over a set of
 adjacent intervals, where the end points of the
 intervals
 are called breakpoints. The coefficients of the polynomials are
 determined by specifying conditions, such as continuity and differentiability,
 at the breakpoints. 
 Additional conditions at a breakpoint can be specified by expanding 
 the sequence of breakpoints into a sequence of 
 {\em knots}, where multiple consecutive knots can have 
 the same breakpoint value.
 
 For a fixed set of $L$ knots $\overline{\tau} = (\tau_0,\tau_1,\ldots,\tau_{L-1})$, 
 the set of all splines defined by $\overline{\tau}$ and 
 having polynomial order $k$ ($=4$ for a cubic polynomial) is a linear vector 
 space of dimensionality $L-k$. The set of B-spline functions, denoted by 
 $\mathcal{B}_{i,k}(t;\overline{\tau})$, $i = 0,1,\ldots,L-k-1$,  constitutes a basis for this space. They can be obtained using the recursion relations~\cite{DEBOOR197250},
 \begin{eqnarray}
  \mathcal{B}_{i,1}(t;\overline{\tau}) & = & \left\{\begin{array}{cc}
      1,  &  \tau_i <= t < \tau_{i+1} \\
      0 & {\rm else}
  \end{array}\right.\;,
  \label{charfunc}
  \\
\mathcal{B}_{i,k}(t;\overline{\tau}) & = & \frac{t - \tau_i}{\tau_{i+k-1} - \tau_i} \mathcal{B}_{i,k-1}(t;\overline{\tau})
              + \nonumber\\
              && \frac{\tau_{i+k} - t}{\tau_{i+k} - \tau_{i+1}} \mathcal{B}_{i+1,k-1}(t;\overline{\tau})\;.
\label{deboor}
 \end{eqnarray}
 From Eq.~(\ref{charfunc}), $B_{i,1}(t)=0$ when $\tau_i = t = \tau_{i+1}$, and any term in 
 Eq.~(\ref{deboor}) that has a zero in the denominator (due to knot multiplicity) will be 
 set to zero by this condition.
 It can be shown that
 $\mathcal{B}_{i,k}(t;\overline{\tau}) = 0$ for $t\notin [\tau_i,\tau_{i+k})$ and positive 
 in the interior of this interval.
 
For  generating B-splines numerically,  we use  routines in the GNU Scientific Library (GSL)\cite{gough2009gnu}. In these routines, 
 the end knots have a multiplicity of $k$ for a spline of order $k$. Thus, the number of B-splines generated is two more than the number of breakpoints. 
 However, since the
 B-splines in this scheme  at the end breakpoints are discontinuous, we 
 always set their corresponding coefficients to zero.
 Therefore, the amplitude envelope spline is a linear combination of $M$ B-splines
 as shown in Eq.~(\ref{aoft_spline}).
 \section{Spectrogram}
 \label{app:spectrogram}
 For a given data sequence $\overline{x}$ of
length $N$,
define the windowed sequence $\overline{x}^L_a$ of length $L<N$ and offset $a$, $\overline{x}^L_a = (x_a,x_{a+1},\ldots,x_{a+L-1})$, 
$0\leq a \leq N-L$.
Then a spectrogram, $\mat{S}^L$, is given by, 
\begin{eqnarray}
[\mat{S}^L]_{mn} & = & \left| \left[\mat{F} (\overline{w}^L.*\overline{x}^L_n)\right]_m\right|^2
\end{eqnarray}
where $\overline{w}^L$ is a window sequence of length $L$,
$m = 0,1,\ldots,\left\lfloor L/2 \right\rfloor$,
and $n = 0, K, 2K, \ldots, \left\lfloor (N-L)/K \right\rfloor-1$ with $1\leq K\leq L$.
Here, $L-K$ specifies the overlap between consecutive windowed sequences. 
In this paper, $\overline{w}^L$ is always a Hamming window and $K = \left\lfloor0.25L\right\rfloor$.
An element at row $i$ and column $j$ of $\mat{S}^L$ is called a pixel, $(i,j)$, and 
$[\mat{S}^L]_{ij}$ is its amplitude.
 
\section{Time-frequency Clustering}
\label{app:clustering}

We present the definition of a time-frequency cluster as used in this paper
as well as the algorithm used for producing clusters. See
Appendix~\ref{app:spectrogram} for the notation used here.

Given a spectrogram $\mat{S}^L$ and a threshold $\eta^L$, define the binary matrix 
$\mat{B}^L$,
\begin{eqnarray}
[\mat{B}^L]_{mn} & = & \left\{
\begin{array}{cc}
    0 & [\mat{S}^L]_{mn} < \eta^L  \\
    1 & [\mat{S}^L]_{mn} \geq \eta^L 
\end{array}
\right.\;,
\end{eqnarray}
One can represent $\mat{B}^L$ as an image with pixels colored black when they have 
amplitude 1 and white otherwise. This has led to the common terminology,  following~\cite{sylvestre2002time},
where a pixel
with amplitude 1 is called a black pixel (BP), $\mat{B}^L$ is called 
the BP map, and $\eta^L$ is called the BP threshold.

Define pixels $(i,j)$ and $(p,q)$ to be nearest neighbors if 
$(p -i,q-j) \in \{-1,0,1\}\times \{-1, 0, 1\}$. We 
call a non-empty sequence of pixels
a {\em path} if it is a sequence of only nearest neighbors, and two pixels
are {\em connected} if they are members of a path.
A non-empty set of 
black pixels is defined to be a cluster if each element of the set
is connected to every element of the set by a path that consists of only the 
elements of the set. 

To distinguish noise and signal induced clusters, we put
a threshold on the cluster integrated power $P_C^L$,
which is
defined as
\begin{eqnarray}
P_C^L & = & \sum_{(i,j)\in C} [\mat{S}^L]_{ij}\;.
\label{eq:clstrintpow}
\end{eqnarray}
For a given data realization and window length $L$,  the union
of pixels from all the clusters for which  
$P_C^L$ exceeds some threshold
is called a time-frequency event, or just an event when there
is no scope for confusion. 

 \section{Clustering based search method}
 \label{app:clusteringsearch}
 
 The steps below describe the clustering based search method used in this paper 
and how it is initialized in our simulations. See Appendix~\ref{app:spectrogram} and \ref{app:clustering} for the notation used here.
\begin{enumerate}
\item Choose a set of values of $L$ to allow multi-resolution analysis.
The frequency spacing between pixels in a column of $\mat{S}^L$ is given by
$f_s/L$~Hz, where $f_s = 4096$~Hz is the sampling frequency  (see Sec.~\ref{sims}).
Following the frequency resolutions used in the 
analysis of GW150914 by the 
Coherent WaveBurst algorithm~\cite{PhysRevLett.116.061102}, we pick
$L = 2^n$, $n = 5,6,\ldots,11$, leading to frequency spacings of
$128, 64, 32, 16, 8, 4$, and  $2$~Hz respectively.
\item Obtain the BP threshold, $\eta^L$, for a target BP rate, $r_{\rm BP}^L$,
in $H_0$ data.
 For the 
 noise model used here, $[\mat{S}^L]_{mn}$
 has an exponential distribution, and  
 assuming that pixels are statistically independent, the 
 BP threshold is given by
\begin{eqnarray}
\eta^L & = & \|\overline{w}^L\|^2\ln \left(\frac{(\left\lfloor L/2 \right\rfloor+1)\left\lfloor (N-L)/K \right\rfloor}{r^L_{\rm BP}\times  N/f_s}\right)\;.
\end{eqnarray}
We set
$r^L_{\rm BP}=1000/7$~BP/sec , leading to an overall rate of $1000$~BP/sec across all
the $7$ window lengths. 
\item Estimate the threshold on cluster integrated power, $P^L_C$. 
We use a target rate
 of $10^{-3}/7$ clusters/sec for each $L$ in $H_0$ data.  
 This results in 
 $\approx 3$ as the expected number of noise induced clusters
 over the entire $2\times 10^4$~sec
 of $H_0$ data for each $L$.
  The corresponding threshold on $P^L_C$ is, therefore,
  taken to be
  the third largest integrated power over all the clusters found for that $L$.
\end{enumerate}
With the thresholds determined as described above, we run the method
on realizations of $H_1$ data for each signal and each ${\rm SNR}$. 

For estimating detection probability, we count events found 
across all the values of $L$ for a single data realization as 1 instance of 
detection. This grouping is an essential part of any multi-resolution analysis
since the same signal can produce clusters across multiple levels of resolution.

Strictly speaking, the same grouping should also be used for clusters obtained from 
$H_0$ data realizations but this is  unnecessary in practice
because the probability of 
clusters appearing across multiple values of $L$ for a single $H_0$ realization, at
the low rate of $10^{-3}/7$~clusters/sec per $L$, is extremely small.
Hence, at low rates of cluster production in $H_0$ data, individual 
 clusters can be identified with instances of detection.
 Thus, the overall rate of $10^{-3}$~clusters/sec that was set above matches
 the FAR of $10^{-3}$~events/sec used for SEECR in Sec.~\ref{seecr_detperf}.

 \bibliography{references.bib,refs_soumya_dod.bib,pta4gw.bib,sdmLasso.bib,mohanty_bib.bib}
\end{document}